\documentclass[%
 reprint,
 amsmath,amssymb,
 aps,
floatfix,
]{revtex4-2}

\usepackage{graphicx}
\usepackage{dcolumn}
\usepackage{bm}

\begin{document}

\preprint{APS/123-QED}

\title{Motility and Phototaxis of \textit{Gonium}, the Simplest Differentiated Colonial Alga}

\author{H\'el\`ene de Maleprade$^{1}$}
 \author{Fr\'ed\'eric Moisy$^{2}$, Takuji Ishikawa$^{3}$}

\author{Raymond E. Goldstein$^{1}$}%
 \email{R.E.Goldstein@damtp.cam.ac.uk}
\affiliation{%
 $^{1}$Department of Applied Mathematics and Theoretical 
Physics, Centre for Mathematical Sciences, University of Cambridge, Wilberforce Road, Cambridge CB3 0WA, 
United Kingdom\\
$^{2}$Laboratoire FAST, Universit\'e Paris-Sud, CNRS, Universti\'e Paris-Saclay - Paris, France\\
$^{3}$Department of Finemechanics, Graduate School of Engineering, Tohoku University, 6-6-01 Aoba, Aramaki, Aoba-ku, Sendai 980-8579, Japan
}%

\date{\today}

\begin{abstract}
Green algae of the \textit{Volvocine} lineage,  
spanning from unicellular
\textit{Chlamydomonas} to vastly larger \textit{Volvox}, are models 
for the study of the evolution of multicellularity, flagellar
dynamics, and developmental processes.  Phototactic steering in
these organisms occurs without a central nervous system, driven solely by the 
response of individual cells.  All such algae
spin about a body-fixed axis as they swim; directional photosensors on 
each cell thus receive periodic signals when that axis is not aligned
with the light.  The flagella of \textit{Chlamydomonas} and \textit{Volvox} 
both exhibit an adaptive response to such signals in a manner that allows for 
accurate phototaxis, but in the former the two flagella have distinct responses, while the
thousands of flagella on the surface of spherical \textit{Volvox} colonies have essentially identical behaviour.
The planar $16$-cell species \textit{Gonium pectorale} thus presents a conundrum, for its central
$4$ cells have a \textit{Chlamydomonas}-like beat that provide propulsion normal to
the plane, while its $12$ peripheral cells generate rotation around the normal through a
\textit{Volvox}-like beat. 
Here, we combine experiment, theory, and computations to reveal how \textit{Gonium}, perhaps the 
simplest differentiated colonial organism, achieves phototaxis.
High-resolution cell tracking, particle image velocimetry of flagellar driven flows, 
and high-speed imaging of flagella on micropipette-held colonies show how, in the 
context of a recently introduced model for \textit{Chlamydomonas} phototaxis, 
an adaptive response of the peripheral cells alone leads to photo-reorientation of the
entire colony. The analysis
also highlights the importance of local variations in flagellar beat dynamics 
within a given colony, which can lead to enhanced reorientation dynamics.
\end{abstract}

\maketitle


\section{\label{sec:Intro}Introduction}

Since the work of A. Weismann on germ-plasm theory in biology \cite{Weismann1892} 
and of J.S. Huxley on the nature of the individual in evolutionary theory \cite{Huxley1912},
the various species of green algae belonging to the 
family \textit{Volvocaceae} have been recognized as important ones in the
study of evolutionary transitions from uni- to multicellular life.  In a modern biological
view \cite{Kirkbook1998}, this significance arises from a number of specific features
of these algae, including the fact that they are an extant family (obviating the need to study
fossils), are readily obtainable in nature, have been studied from a variety of 
perspectives (biochemical, developmental, genetic), and have had significant ecological studies.
From a fluid dynamical perspective \cite{Goldstein2015ARFM}, their relatively large size and
easy culturing conditions allow for precise studies of their motility, the flows they
create with their flagella, and interactions between organisms, while their high degree of symmetry simplifies theoretical descriptions of those same phenomena \cite{Goldstein2016JFM}.

As they are photosynthetic, the ability of these algae to execute phototaxis is
central to their life.  Because the lineage spans from unicellular to large colonial forms, it can be
used to study the evolution of multicellular coordination of motility.
Motility and phototaxis of motile green algae have been the subjects of an extensive literature in recent 
years \cite{hoops1997motility, drescher2010fidelity, drescher2010direct, guasto2010oscillatory, bennett2015steering, arrieta2017phototaxis, tsang2018polygonal, leptos2018adaptive}, focusing primarily on the two extreme cases: unicellular {\it Chlamydomonas} and much larger {\it Volvox}, with species 
composed of $1,000-50,000$ cells. \textit{Chlamydomonas}, the simplest member of the \textit{Volvocine} family, 
swims typically by actuation of its two flagella in a breast stroke, combining propulsion and slow 
body rotation. It possesses an \textit{eye spot}, a small area highly sensitive to light 
\cite{foster1980light, hegemann2008algal}, which triggers the two flagella differently \cite{kamiya1984submicromolar}. Those responses are adaptive, on a timescale matched to the 
rotational period of the cell body \cite{Josef2005, Josef2006, Yoshimura2001}, and allow cells to scan 
the environment and swim towards light \cite{leptos2018adaptive}. Multicellular \textit{Volvox} shows a higher level of complexity, with differentiation between interior germ cells 
and somatic cells dedicated to propulsion.  Despite lacking a central nervous system to coordinate its cells, {\it Volvox} exhibits accurate phototaxis.  This is also achieved by an adaptive response to changing
light levels, with a response time tuned to the \textit{colony} rotation period which creates a differential 
response between the light and dark sides of the spheroid \cite{kirk2004volvox,drescher2010fidelity}.

\begin{figure}
\includegraphics[width = 8cm]{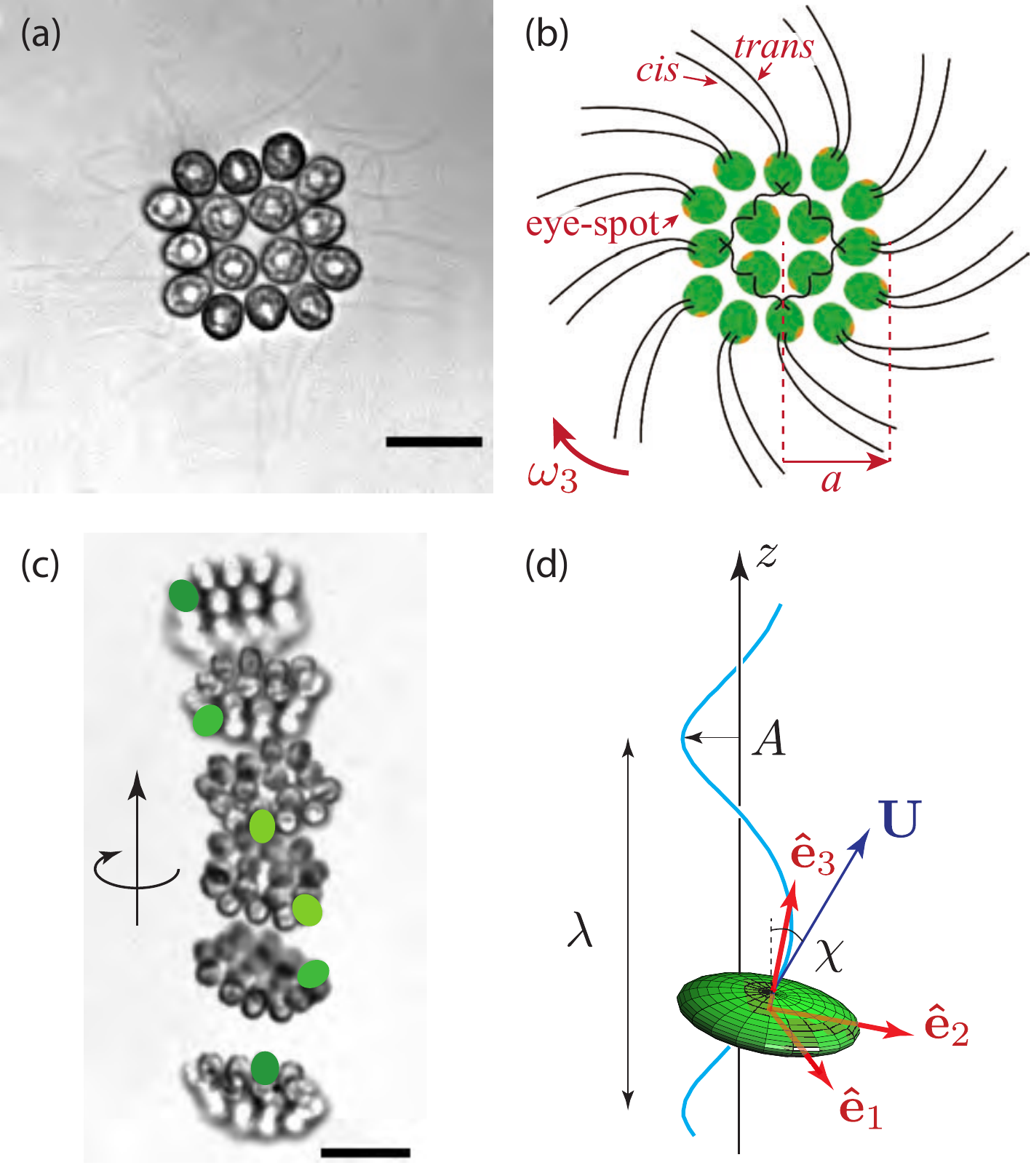}
\caption{\label{fig:GoniumPic}Geometry and locomotion of \textit{Gonium pectorale}. 
(a) $16$-cell colony. Each cell has two flagella, $30-40$ $\mu$m long. Scale bar is $10\ \mathrm{\mu m}$.  
(b) Schematic of a colony of radius $a$: $16$ cells (green) each with one eye spot (orange dot). 
The \textit{cis} flagellum 
is closest to the eye-spot, the \textit{trans} flagellum is furthest \cite{coleman2012}. Flagella of the central cells beat in 
an opposing breaststoke, while the peripheral flagella beat in parallel. The pinwheel organization of the 
peripheral 
flagella leads to a left-handed body rotation at a rate $\omega_3$. 
(c) Upward swimming of a colony. Superimposition of images separated by $0.4$ s. Green spots label one 
specific cell 
to highlight the left-handed rotation. Scale bar is $20$ $\mu$m.
(d) Sketch of a helical trajectory: a colony (green ellipsoid) swims with velocity \textbf{U} along 
an oscillatory path 
(blue line) of wavelength $\lambda$, amplitude $A$ and pitch angle $\chi$. The frame  
$({\bf \hat e}_1, {\bf \hat e}_2, {\bf \hat e}_3)$ is attached to the \textit{Gonium} body.}
\end{figure}

In light of the above, a natural questions is: how does the 
simplest \textit{differentiated}  organism achieve phototaxis?  In the
Volvocine lineage the species of interest is \textit{Gonium}. This $8$- or $16$-cell 
colony represents one of the first steps to true multicellularity \cite{arakaki2013}, 
presumed to have evolved from the unicellular common ancestor earlier than other Volvocine algae \cite{herron2008}. It is also the first to show cell 
differentiation. We focus here on $16$-cell colonies, which show a higher degree of symmetry than those 
with $8$, but our results apply to both.

A 16-cell {\it Gonium} colony is shown in Fig.~\ref{fig:GoniumPic}. It is organized into two concentric 
squares of respectively $4$ and $12$ cells, each biflagellated, held together by an extra-cellular matrix~\cite{nozaki1990ultrastructure}. All flagella point out on the same side: it exhibits a much lower 
symmetry than \textit{Volvox}, lacking anterior-posterior symmetry. Yet it performs similar functions 
to its unicellular and large colonies counterparts as it mixes propulsion and body rotation, and swims efficiently towards light \cite{moore1916mechanism, mast1916process, hoops1997motility}. 
The flagellar organization of inner and peripheral cells deeply differs \cite{greuel1985, harper1912}: 
central cells are similar to \textit{Chlamydomonas}, with the two flagella beating in an opposing breast 
stroke, and contribute mostly to the forward propulsion of the colony. Cells at the periphery, however, 
have flagella beating in parallel, in a fashion close to \textit{Volvox} cells \cite{coleman2012}. 
This minimizes steric interactions and avoids flagella crossing each other \cite{hoops1997motility}. 
Moreover, these flagella are implanted with a slight angle and organized in a pin-wheel fashion 
(see Fig.~\ref{fig:GoniumPic}b) \cite{greuel1985}: their beating induces a left-handed rotation 
of the colony, highlighted in Figs.~\ref{fig:GoniumPic}c\&d and in Supplementary Movie 1. Therefore, 
the flagella structure of \textit{Gonium} reinforces its key position as intermediate in the evolution 
towards multicellularity and cell differentiation.

These small flat assemblies show intriguing swimming along helical trajectories - with their body plane 
almost normal to the swimming direction - that have attracted the attention of naturalists since the 
eighteenth century~\cite{muller1782,mast1916process,moore1916mechanism}.
Yet, the way in which {\it Gonium} colonies bias their swimming towards the light remains unclear. 
Early microscopic observations have identified differential flagellar activity between the illuminated 
and the shaded sides of the colony as the source of phototactic
reorientation~\cite{mast1916process,moore1916mechanism}. Yet, a full fluid-dynamics description, 
quantitatively linking the flagellar response to light variations and the hydrodynamic forces and 
torques acting on the colony, is still lacking. 
From an evolutionary perspective, phototaxis in {\it Gonium} also raises a number of fundamental issues: to 
what extent is the phototactic strategy of the unicellular ancestor retained in the colonial form? How is 
the phototactic flagella reaction adapted to the geometry and symmetry of the colony, and how does it 
lead to an effective reorientation?

Taking this specific structure into account, here we aim to understanding how the individual cell 
reaction to light leads to reorientation of the whole colony.  Any description of phototaxis must build on 
an understanding of unbiased swimming, so we first focus on the helical 
swimming of \textit{Gonium}, and show how it results from an uneven distribution of forces around the colony. 
We then investigate experimentally its phototaxis, by describing the reorientation trajectories and 
characterising the cells' response to light.  That response is shown to be adaptive, and we therefore 
extend a previously introduced model for such a response to the geometry of \textit{Gonium}, and show 
how the characteristic relaxation times are finely tuned to the {\it Gonium} body shape and rotation 
rate to perform efficient phototaxis.

\section{\label{sec:Swimming}Free Swimming of \textit{Gonium}}

\subsection{Experimental observations}

\begin{figure}
\includegraphics[width = 8cm]{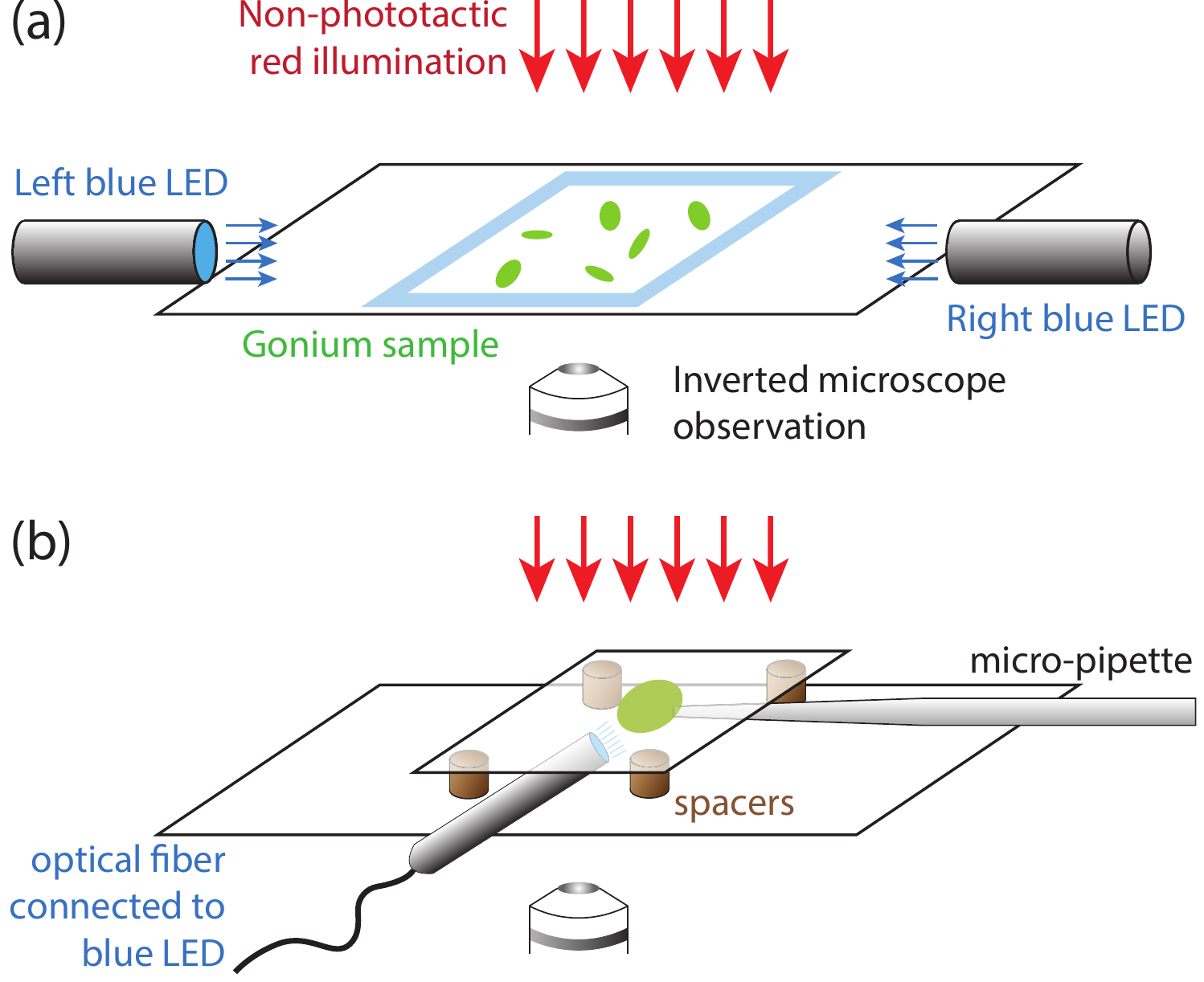}
\caption{\label{fig:exp_setup} Phototaxis experiments. (a) \textit{Gonium} colonies swim in a sealed chamber 
made of two glass slides, with non-phototactic red illumination from above, on the stage of an inverted microscope 
connected to a high speed video camera. Two blue LEDs on the right- and left-hand sides of the chamber can 
independently shine light with controllable intensities. 
(b) Micropipette experiments.  A micropipette of inner diameter $\sim 20$ $\mu$m holds a colony (green disc) 
in a chamber made of two glass slides, spaced to allow room for optical fiber connected to a blue LED to 
enter the chamber.}
\end{figure}

\begin{figure*}
\includegraphics[width = 17cm]{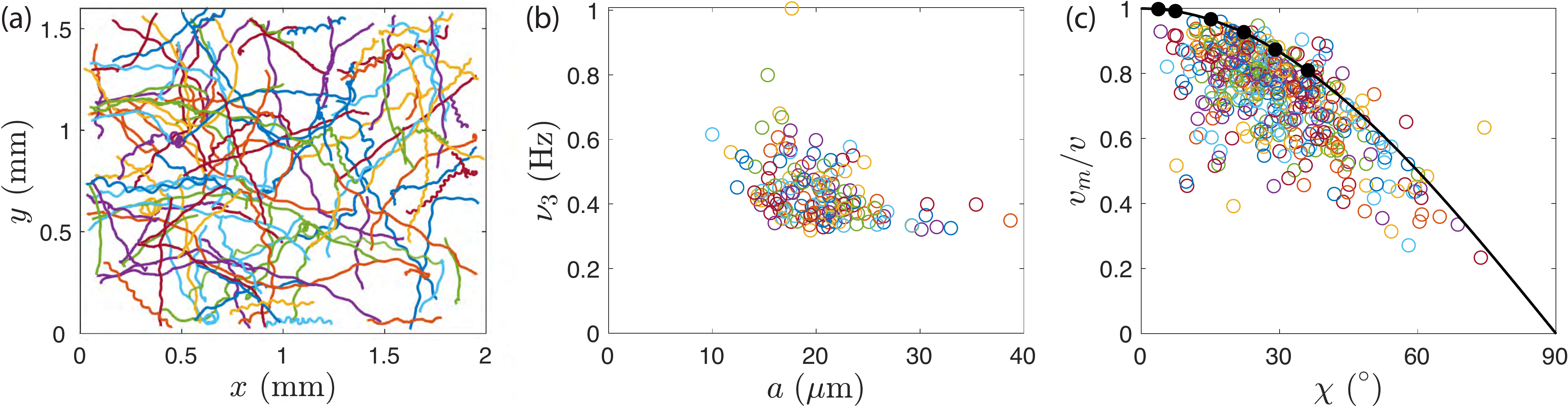}
\caption{\label{fig:GpSwimming} Free swimming of \textit{Gonium}. (a) Trajectories of many colonies under 
non-phototactic illumination, showing random swimming. Each colored line shows the path of one colony 
(sample size: $117$ colonies).
(b) Body rotation frequency $\nu_3$ as a function of colony radius $a$ (sample size: $159$ colonies). 
(c) Mean velocity $v_m$ normalised by the instantaneous swimming velocity $v$ as a function of helix pitch angle 
$\chi$ (sample size: $430$ colonies). The black line indicates relation $v_m/v = \cos{\chi}$ and black discs along 
it are results from the numerical simulations as described in text.}
\end{figure*}

We recorded trajectories of \textit{Gonium} colonies freely swimming in a sealed chamber on an inverted microscope, 
connected to a high speed video camera, as sketched in Fig. \ref{fig:exp_setup}a and detailed in 
Appendix \ref{sec:app:Materials}.  To obtain unbiased random swimming trajectories, we used red light illumination.
Trajectories were reconstructed using a standard tracking algorithm, and shown in Fig. \ref{fig:GpSwimming}a and Supplementary Movie 2. They exhibit 
a large variation in waviness, with some colonies swimming along nearly straight lines, while others show highly curved 
helices. From Fig. \ref{fig:GoniumPic}c-d, we infer that a colony performs a full body rotation per helix wavelength.
This observation suggests that the waviness of the trajectories arises from an uneven distribution of forces developed 
by the peripheral flagella, with the most active flagella located on the outer side of the helix.


From image analysis, we extracted for each colony the body radius $a$, the body rotation frequency $\nu_3 = \omega_3/2\pi$, 
the instantaneous swimming velocity $v$ projected in the plane of observation, and the mean swimming velocity $v_m$. 
The rotation frequency $\nu_3 \sim0.4$ Hz, decreases slightly with colony size as shown in Fig. \ref{fig:GpSwimming}b. 
This places \textit{Gonium} in a consistent intermediate position between \textit{Chlamydomonas} and \textit{Volvox}, 
whose radii are respectively about $5$ $\mu$m and $200$ $\mu$m, and whose rotation rates are $2$ Hz and  $0.2$ Hz
\cite{witman1993, drescher2010fidelity}.  With a typical flagellar beating frequency of $10$ Hz (see measurements in Sec. \ref{sec:Phototaxis}), 
there are 
$\sim 25$ flagella strokes per body rotation, a value similar to that in \textit{Chlamydomonas}.

There is a significant correlation between the swimming velocity and the waviness of the trajectories, displayed in Fig. \ref{fig:GpSwimming}c. We quantify the degree of waviness with the pitch angle $\chi$, defined as the angle between the swimming speed $\textbf{U}$ and the helix axis ${\bf \hat e}_z$, such that $\tan \chi= (2\pi A/\lambda)$, with $A$ and $\lambda$ the helix amplitude and wavelength (see Fig. \ref{fig:GoniumPic}d). 
Swimming in helices is clearly at the cost of the swimming efficiency: the ratio $v_m/v$ shows a marked decrease with the angle $\chi$. The data is well described by the simple law $v_m/v = \cos{\chi}$ expected for a helix traced out at constant instantaneous velocity (black line in Fig. \ref{fig:GpSwimming}c). 
This geometrical law is valid for a $3$-dimensional velocity $v_{3D}$, while we only have access to its two-dimensional projection. In Appendix \ref{app:v3D} we show that $v$ (the projected velocity) provides a reasonable approximation to the real velocity $v_{3D}$ for $\chi$ not too large. 
The average pitch angle in Fig. \ref{fig:GpSwimming}c is $\chi \simeq 30^\circ \pm 13^\circ$, corresponding to a mean velocity experimentally $30\%$ slower than the instantaneous velocity.  In spite of this decreased swimming efficiency, the surprisingly high level of waviness found in most {\it Gonium} colonies suggest that this trait provides an evolutionary advantage.

\subsection{\label{sec:Kin}Fluid dynamics of the swimming of {\it Gonium}}


We introduce here the fluid dynamical and computational description of 
the swimming of \textit{Gonium}.
With a typical size of 40~$\mu$m and swimming speed of 40~$\mu$m/s, the Reynolds number is of order $10^{-3}$, so the swimming is governed by the Stokes equation, \textit{i.e.} by the balance between the force and torque induced by the flagella motion to those arising from viscous drag \cite{lauga2009hydrodynamics}.

\begin{figure}
\includegraphics[width = 6.5cm]{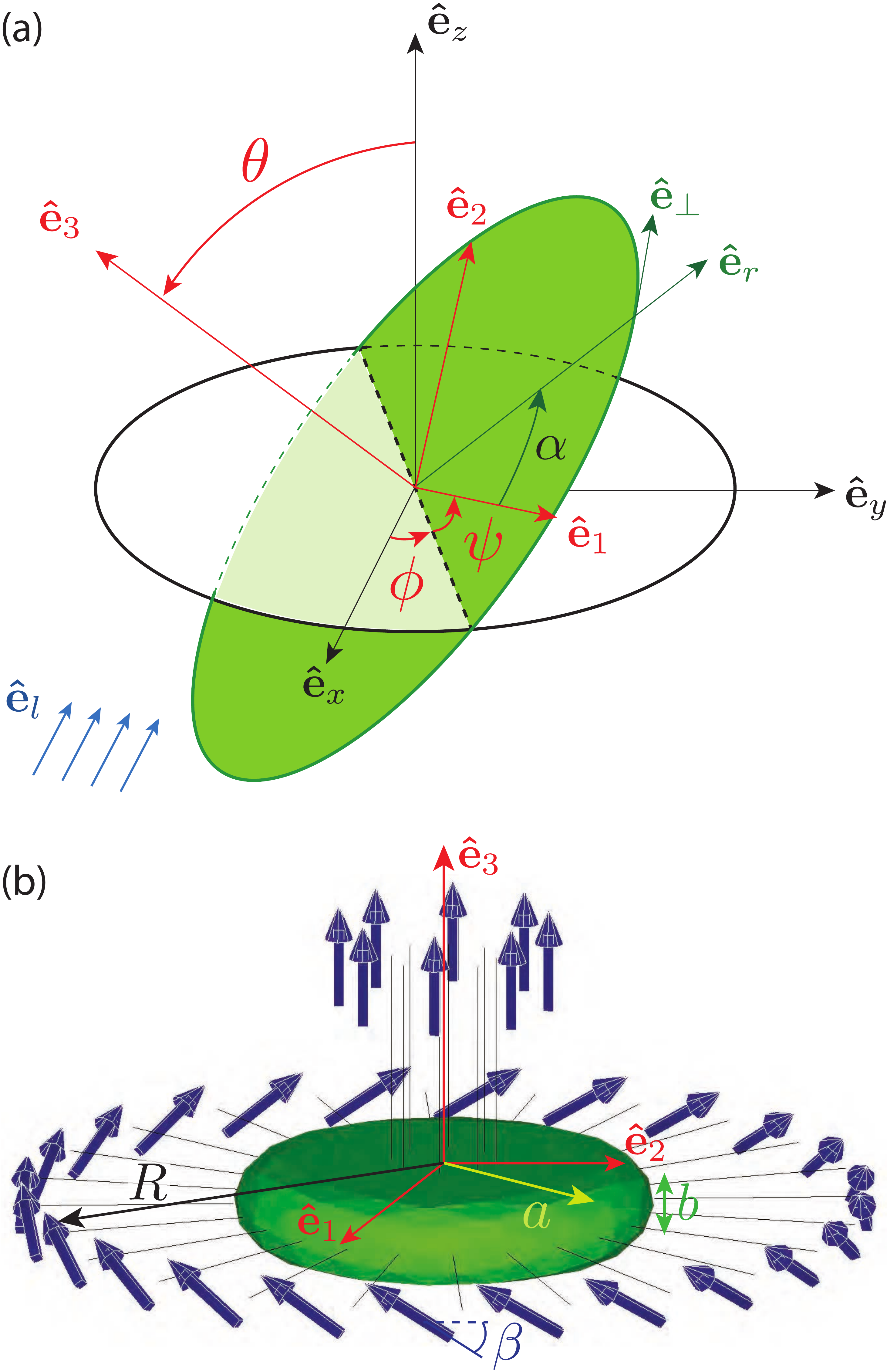}
\caption{\label{fig:geometry} Details of computational geometry.
(a) Coordinates system and Euler angles. The frame $({\bf \hat e}_x, {\bf \hat e}_y, {\bf \hat e}_z)$ is attached to 
the laboratory; the frame $({\bf \hat e}_1, {\bf \hat e}_2, {\bf \hat e}_3)$ is attached to the \textit{Gonium} 
body (green disk), with ${\bf \hat e}_3$ the symmetry axis and $({\bf \hat e}_1, {\bf \hat e}_2)$ in the body plane. 
Euler angles $(\theta, \phi, \psi)$ relate the two frames: by definition, $\theta$ is the angle between ${\bf \hat e}_z$
and ${\bf \hat e}_3$, $\phi$ is the angle from  ${\bf \hat e}_x$ to the line of nodes (dotted line), and $\psi$ is the 
angle from the line of nodes to ${\bf \hat e}_1$.
In the \textit{Gonium} body plane $({\bf \hat e}_1, {\bf \hat e}_2)$, the flagella are labeled by the angle 
$\alpha$, with $({\bf \hat e}_r, {\bf \hat e}_\perp)$ the corresponding local frame such that 
$\cos \alpha = {\bf \hat e}_1 \cdot {\bf \hat e}_r$. For the computation of the phototactic response, we assume 
the light is incident along ${\bf \hat e}_l = -{\bf \hat e}_x$ (blue arrows).
(b) \textit{Gonium} geometry for simulations. The body (in green) is a thick disc with radius $a = 20~\mathrm{\mu m}$ 
and thickness $b=8~\mathrm{\mu m}$. Flagella (length $20~\mathrm{\mu m}$) associated with a point force $20~\mathrm{\mu m}$ 
away from the cell body are attached to the body. The $8$ central flagella generate thrust while the $24$ peripheral 
ones are tilted by $\beta \simeq 30 ^\circ$, and generate both thrust and rotation.}
\end{figure}

We model a colony as a thick disk of symmetry axis ${\bf \hat e}_3$. This disk includes the averaged cell body radius $a$, and also the flagella, which contribute significantly to the total friction. We therefore consider an effective radius $R$ encompassing the whole structure. In the frame $({\bf \hat e}_1, {\bf \hat e}_2, {\bf \hat e}_3)$ attached to the body, the viscous force ${\bf F}_v$ and torque ${\bf L}_v$ are linearly related to the velocity ${\bf U}$ and angular velocity ${\bf \Omega}$ through
\begin{subequations}
\begin{align}
{\bf F}_v &= -  \eta R
\begin{bmatrix}
    k_1 & 0 & 0 \\
    0 & k_1 & 0 \\
    0 & 0 & k_3 \\ 
\end{bmatrix}
{\bf U},\\
{\bf L}_v &= - \eta R^3
\begin{bmatrix}
    l_1 & 0 & 0 \\
    0 & l_1 & 0 \\
    0 & 0 & l_3 \\ 
\end{bmatrix}
{\bf \Omega},
\label{eq:lv}
\end{align}
\end{subequations}
where $\eta = 1$~mPa~s is the viscosity of water, and the numbers $(k_1, k_3, l_1, l_3)$ quantifying the translation and rotation friction along the transverse and axial directions respectively, characteristic of the \textit{Gonium} geometry. Left-handed body rotation implies ${\bf U} \cdot {\bf \Omega} < 0$. The angular velocity ${\bf \Omega}$ is expressed in the body frame by introducing the Euler angles $(\theta, \phi ,\psi)$ defined in Fig. \ref{fig:geometry}a \cite{symon1971chapter},
\begin{equation}
{\bf \Omega} =
\begin{bmatrix}
\dot \theta \cos \psi + \dot \phi \sin \theta \sin \psi \\
-\dot \theta \sin \psi + \dot \phi \sin \theta \cos \psi \\
\dot \psi + \dot \phi \cos \theta
\end{bmatrix}.
\end{equation}

These viscous forces and torques are balanced by the thrust and spin induced by the action of the flagella. The central flagella produce a net thrust $F_c\, {\bf \hat e}_3$ and no torque. The peripheral flagella contribute both to propulsion and rotation: we model them in the continuum limit as an angular density of force in the form $d{\bf f}_p = {\bf f}_p d\alpha$, with ${\bf f}_p(\alpha)= f_{p \parallel} {\bf \hat e}_3 + f_{p \perp} {\bf \hat e}_\perp$, where  $\alpha$ is the angle labeling the flagella and  ${\bf \hat e}_\perp = - \sin \alpha {\bf \hat e}_1 + \cos \alpha {\bf \hat e}_2$ the unit azimuthal vector along the \textit{Gonium} periphery (Fig. \ref{fig:geometry}a). Here, we choose $f_{p \parallel}>0$ and $f_{p \perp}<0$ to ensure a left-handed body rotation, with the ratio $f_{p \parallel}/f_{p \perp} = - \tan \beta$ which we assume independent of $\alpha$, and $\beta$ the tilt angle made by the peripheral flagella (Fig. \ref{fig:geometry}b). The flagella therefore produce a net force and a net torque
\begin{equation}
{\bf F} = F_c\,{\bf \hat e}_3 + \int_0^{2\pi}\!\! {\bf f}_{p} d\alpha, \qquad {\bf L} = \int_0^{2\pi}\!\! {\bf r} \times {\bf f}_{p} d\alpha.
\label{eq:lp0}
\end{equation}
In the absence of phototactic cues and for a perfectly symmetric \textit{Gonium}, ${\bf f}_{p}$ has no $\alpha$ dependence, and the force and torque are purely axial,
\begin{equation}
{\bf F}^{(0)} = \left[F_c + 2\pi f_{p \parallel}^{(0)}\right] {\bf \hat e}_3, \qquad {\bf L}^{(0)} =  2\pi R f_{p \perp}^{(0)} {\bf \hat e}_3,
\end{equation} 
which satisfies ${\bf F}^{(0)} \cdot {\bf L}^{(0)} < 0$. For such a {\it straight swimmer}, the velocity ${\bf U}$ and angular velocity ${\bf \Omega}$ resulting from the balance with the frictional forces and torques are also purely axial, along ${\bf \hat e}_3$ and $-{\bf \hat e}_3$, respectively.

The helical trajectories observed experimentally indicate that in most {\it Gonium} colonies the forces ${\bf f}_{p}(\alpha)$ produced by the peripheral flagella are not perfectly balanced.  Such an unbalanced distribution produces nonzero components of ${\bf F}$ and ${\bf L}$ normal to ${\bf \hat e}_3$, which deflects the swimming direction of the colony. The simplest imbalance compatible with the observed helical trajectories is a modulation of the force developed by the peripheral flagella in the form
\begin{equation}
{\bf f}_{p}(\alpha) ={\bf f}_{p}^{(0)} (1+\xi \cos \alpha),
\label{eq:def}
\end{equation}
with $0\le \xi \leq 1$ a parameter. With this choice, there is a stronger flagellar force along ${\bf \hat e}_1$ and the net force and torque acquire components in the plane $({\bf \hat e}_2, {\bf \hat e}_3)$. Interestingly, the combined effect of these transverse force and torque components is to enable such unbalanced colonies to maintain their overall swimming direction.

To relate the uneven distribution of forces (\ref{eq:def}) to the helical pitch angle $\chi$, we consider the geometry sketched in Fig. \ref{fig:GoniumPic}d: a {\it Gonium} colony swims along a helix whose axis is along ${\bf \hat e}_z$, with ${\bf \hat e}_3$ describing a cone around ${\bf \hat e}_z$ of constant apex angle $\zeta$. The angular velocity vector ${\bf \Omega}$ is therefore along ${\bf \hat e}_z$.
In terms of Euler angles (Fig. \ref{fig:geometry}a), this choice implies $\theta=\zeta$, $\psi = 0$ and ${\bf \Omega} = \dot \phi (\sin \zeta {\bf \hat e}_2 + \cos \zeta {\bf \hat e}_3)$ with constant $\dot \phi<0$ (See Appendix \ref{app:wavy}). 

The helix pitch angle $\chi$ is the angle between ${\bf U}$ and ${\bf \hat e}_z$. Because of the anisotropic resistance matrices and the axial force $F_c$ produced by the central flagella, $\chi$ is larger than the angle $\zeta$ between ${\bf \hat e}_3$ and ${\bf \hat e}_z$: the colony exhibits large lateral excursions while maintaining its symmetry axis ${\bf \hat e}_3$ nearly aligned with the mean swimming direction ${\bf \hat e}_z$. Expanding to first order in the
imbalance amplitude $\xi$, we find from force and torque balance
\begin{equation}
\tan \chi = \frac {\xi}{2} \left( \frac{l_3}{l_1} \tan \beta + \frac{k_3}{k_1} \frac{1}{\tan \beta} \frac{1}{1+F_c/(2\pi f_{p \parallel}^{(0)})} \right),
\label{eq:th0}
\end{equation}
so a symmetric \textit{Gonium} ($\xi=0$) swims straight ($\chi=0$).

\subsection{Comparison with experiments and computations}

The continuous angular density of forces considered in the previous section is convenient to obtain a simple analytical description of the helical trajectories. However, to estimate the main hydrodynamic properties (flagella force and resistance matrices) from measurements, a more realistic description is needed that considers the drag exerted on the cell body by the flow induced by the flagellar forces.
We assume for simplicity that all flagella produce equal individual forces $F_i$. With $4$ central and $12$ peripheral bi-flagellated cells, $F_c = 8 F_i$ for the central cells, and $2 \pi f_{p \parallel}^{(0)} = 24 F_i \sin \beta$ and $2 \pi f_{p \perp}^{(0)} = - 24 F_i \cos \beta$ for the axial and azimuthal force developed by the peripheral cells. We first restrict 
our attention to {\it straight} swimmers with balanced peripheral flagella force ($\xi=0$).

To identify the physical parameters (flagella force $F_i$, flagella tilt angle $\beta$, effective radius $R$, translation and rotation friction coefficients $k_1, k_3, l_1, l_3$) from the experimental observations (swimming speed $v \simeq 40~\mu$m/s and body rotation frequency $\nu_3 \simeq 0.4$~Hz), we combined computations and PIV measurements of the flow around a swimming {\it Gonium} colony. The computations, based on the boundary element method \cite{Itoh2019}, are detailed in Appendix \ref{sec:app:Num}. We use the simplified geometry of Fig. \ref{fig:geometry}b: the cell body is a thick disk of radius $a$ and thickness $b$, with $32$ straight filaments representing the flagella pairs from the $16$ cells, and a set of $32$ point forces capturing the result of the flagella beating.   These filaments also contribute to the hydrodynamic drag, and hence to the values of the friction coefficients $k_1, k_3, l_1, l_3$. Fagella lengths are $30-40~\mu$m, but we consider in the computations the average of their shape over a beat cycle as contributing to the drag, which we expect to be in the range of 
$20-30$ $\mu$m. Comparing the velocity fields from PIV and numerics (Figs. 5a,b), we identify the location of the point force at about 20~$\mu$m from the cell body. 

\begin{figure}
\includegraphics[width = 8.5cm]{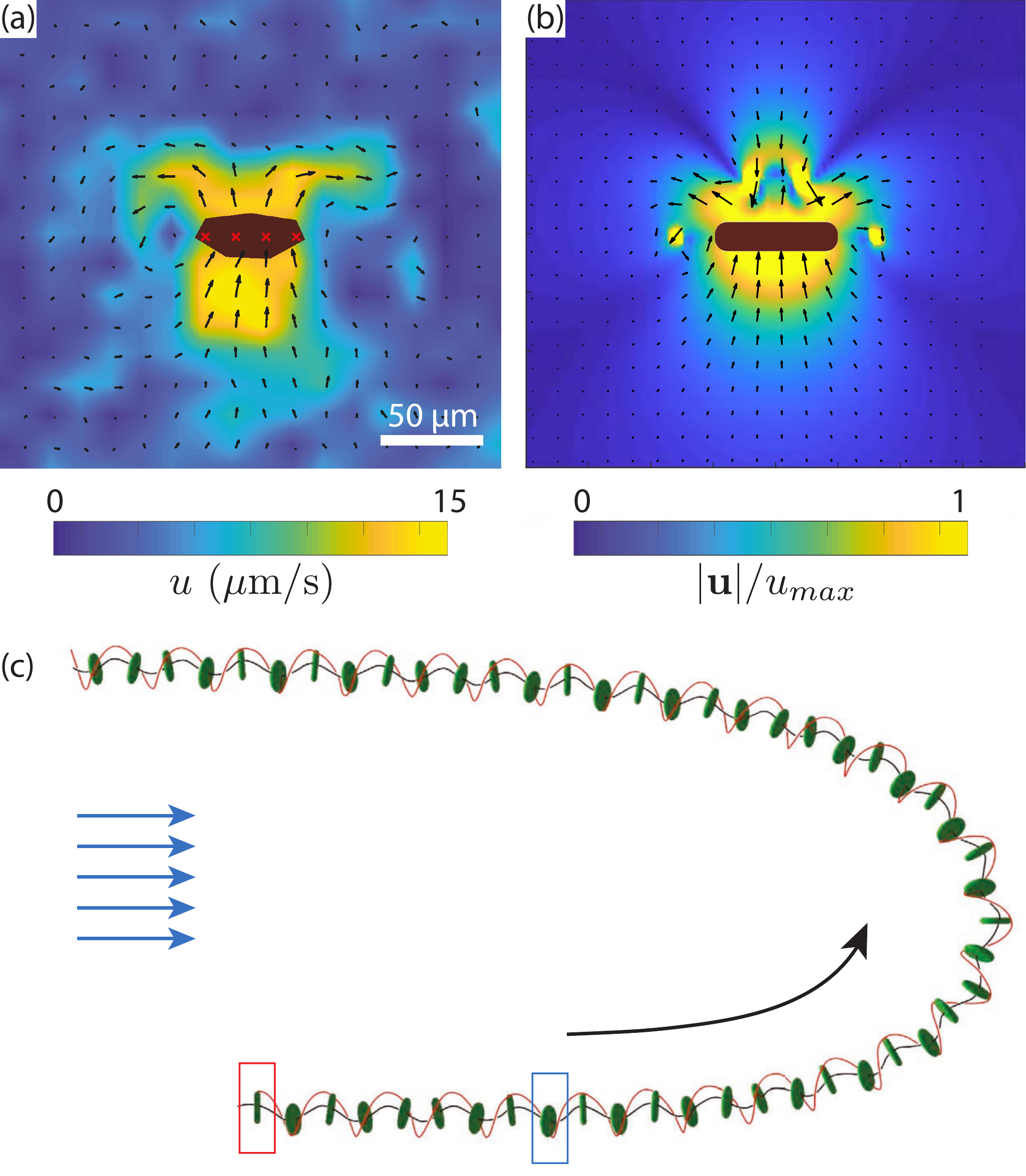}
\caption{\label{fig:swim} Photoresponse of \textit{Gonium}. (a-b) Flow fields around a colony swimming 
towards the top of the image, in the laboratory frame. 
(a) Micro-PIV measurements (the body is in black with red crosses). The background color shows the norm of the 
velocity, varying here up to 15~$\mu$m/s.  Scale bar is $50$ $\mu$m. 
(b) Numerical flow field using the point-force model.
(c) Wavy trajectory of a colony, numerically computed for $\xi = 0.4$ and $\mu s_0 = 0.5$. The initial position 
is shown by the red rectangle, and the time when light is switched on from the left is indicated by the blue rectangle. 
Black line shows the trajectory, and the red line follows the position of the point of maximal force, $\alpha = 0$, 
highlighting rotation of the body.}
\end{figure}

From this geometry, we compute the friction coefficients $k_1, k_3, l_1, l_3$ for several flagella lengths between $20$ and $30~\mu$m, and tune the intensity of the point forces $F_i$ and tilt angle $\beta$ to obtain a good match to the experimental swimming and angular speeds. 
Tuning the flagella length to $20~\mu$m, so that $R \simeq 2a \simeq 40~\mu$m, appeared as the best match between experimental observations and numerical results.
This corresponds to a point force per flagella $F_i \simeq 3.5$~pN and a tilt angle $\beta \simeq 30^\circ$. The friction coefficients are $k_1 \simeq 10, \  k_3 \simeq 13, \  l_1 \simeq 6$, and $l_3 \simeq 8$, as detailed in Appendix \ref{sec:app:Num}. 
The flow field computed from these parameters (Fig. \ref{fig:swim}b) shows an overall structure in reasonable agreement with the PIV measurement performed around a freely swimming {\it Gonium} colony, which validates the methods. As expected, the far-field structure is typical of a {\it puller} swimmer, with inward flow in the swimming direction and outward flow normal to it.

\begin{figure*}[t]
\includegraphics[width = 17cm]{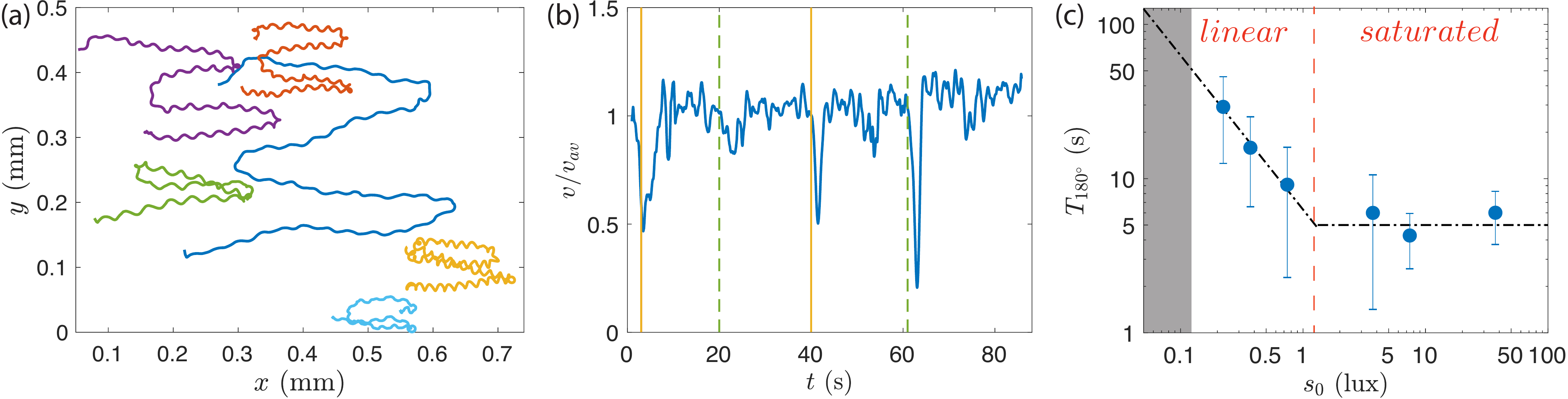}
\caption{\label{fig:PhototaxEvidence}
Phototactic reorientation of \textit{Gonium}. (a) Reorientation trajectories of colonies under blue light 
stimulation shone alternatively from right to left twice during $20$ s.
Each color line corresponds to a single colony and shows a swimming direction alternating to the right and left 
according to the change in light source position. 
(b) Instantaneous normalized velocity as a function of time for the blue trajectory in (a). Vertical lines indicate a 
change in the light source position: yellow shows times when the light is shone from the right, while green dashed 
lines stand for light coming from the left. 
(c) Reorientation time $T_{180^\circ}$ as a function of the light intensity $s_0$. The line shows 
$T_{180^\circ} \simeq 1/s_0$ for $s_0 < 1$~lux, consistent with Eq.~(\ref{eq:tau}), and a saturation at $T_{180^\circ} = 5~\mathrm{s}$ at larger $s_0$. The grey shaded area corresponds to times longer than the video-camera trigger.}
\end{figure*}

Note that the computed force is the flagellar thrust only which does not take into account the drag force, so that it overestimates the total force exerted by a flagellum on the fluid.
An estimate for the true force is obtained by balancing the viscous force experienced by a colony $F_v = -k_3 \eta R v$ to the propulsive contributions of the flagella $8F_{i} + 24 F_{i} \sin{\beta}
\approx 20 F_i$. The resulting force that a flagellum effectively applies on the fluid is of the order of $1$ pN, a value close to the estimate for \textit{Chlamydomonas} over a beating cycle \cite{guasto2010oscillatory}.



We finally consider helical trajectories produced by unbalanced swimmers ($\xi\ne 0$), and include in the computation angular modulation of the peripheral flagella force (\ref{eq:def}). A typical trajectory is illustrated in Fig. \ref{fig:swim}c and Supplementary Movie 3 (note the non-phototactic part of the trajectory, between the red and blue rectangles). The black and red lines show the helical body trajectory and the position of the maximal force (unit vector ${\bf \hat e}_1$). As expected, the anisotropic resistance matrices characteristic of the {\it Gonium} geometry produce a helical trajectory with large lateral excursion but moderate tilt of the body normal axis.

To relate the helical pitch angle $\chi$ to the parameter $\xi$, we extract from the simulated trajectories the helix amplitude $A$, wavelength $\lambda$ and mean velocity $v_m$, as described in the Appendix \ref{sec:app:Num}. We deduce a pitch angle $\chi$, observed to linearly increase with $\xi$, in close agreement with Eq.~(\ref{eq:th0}). The decrease in swimming efficiency with $\chi$ is also well reproduced, with numerical data closely following the geometrical prediction of $v_m/v = \cos \chi$, as evidenced by the black dots in Fig. \ref{fig:GpSwimming}c. From the experimental measurement $\chi \simeq 30^\circ \pm 13^\circ$, we deduce an average amplitude $\xi \approx 0.4$: the strongest flagella typically produce a force at least twice larger than the weakest ones. This surprisingly large value suggests that waviness is an important feature for the swimming of {\it Gonium}. 

\section{\label{sec:Phototaxis}Phototactic swimming}

\subsection{Experimental observations}

We now turn to the phototactic response of {\it Gonium}, which is triggered by adding to the previous experimental setup two blue LEDs on the sides of the chamber. The simplest configuration has the two lights arranged facing each other, as in Fig. \ref{fig:exp_setup}a.  Figure \ref{fig:PhototaxEvidence}a displays the trajectories of a set of colonies reorienting as the two lights are alternately switched on and off for approximately $20 $s twice in a row, also seen in Supplementary Movie 4. These trajectories show various degrees of waviness, as in the non-phototactic experiments (Fig. \ref{fig:GpSwimming}a), however here their direction is no longer random but rather aligned with the incident light. At each change of light direction, a marked slowdown is also observed, as illustrated in Fig. \ref{fig:PhototaxEvidence}b: just after the change in light, the swimming speed decreases by half for a few seconds, indicating a reduction in the flagella activity \cite{moore1916mechanism}. The variability in this drop can originate from out of plane swimming during reorientation, as we only have access to ($x,y$) projections of the trajectories.

A key feature to model the dynamics of the body reorientation towards light is the dependence of the flagella force on light intensity. We measured the time $T_{180^\circ}$ to perform a turn-over as a function of the light intensity $s_0$ (Fig.~\ref{fig:PhototaxEvidence}c), restricting measurements to $s_0 < 100$~lux, so as to observe only {\it positive} phototaxis; at larger $s_0$, an increasing fraction of colonies display {\it negative} phototaxis, swimming away from the light (the critical light intensity between positive/negative phototaxis varies with time during the diurnal cycle).

\begin{figure*}
\includegraphics[width = 17cm]{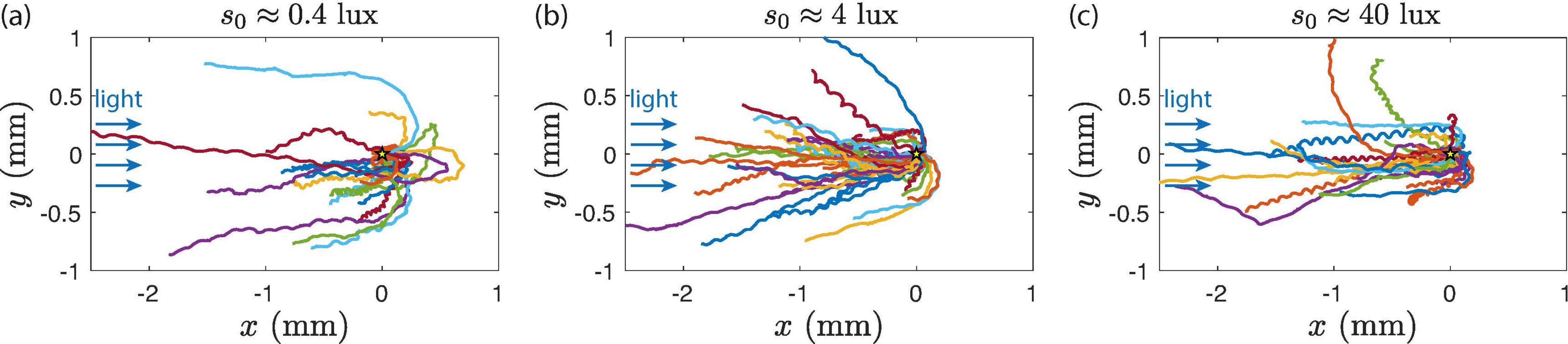}
\caption{\label{fig:TurnOver-compTh} Phototactic turn-over after a change of light incidence. \textit{Gonium} colonies are initially swimming ($t < 0$~s) toward a light of constant intensity on the right. At 
$t = 0$~s, this light is switched off while another of controlled adjustable intensity $s_0$ is shone from 
the left. Trajectories are reported for $t > 0$~s, and have been shifted to the origin at $t = 0$~s.}
\end{figure*}

Figure \ref{fig:PhototaxEvidence}c shows two phototactic regimes: A {\it linear} regime at moderate intensity ($s_0<1$~lux), for which $T_{180^\circ} \sim 1/s_0$, and a {\it saturated} regime at larger $s_0$, for which $T_{180^\circ} \simeq 5~\mathrm{s} \pm 1~$s. In what 
follows we focus on the linear regime, in which the flagella activity is proportional to the light intensity. The constant reorientation time in the saturated regime may originate from a biological saturation in the signal transmission from the eye-spot to the flagella, or from a hydrodynamic limitation in the reorientation process itself: In this saturation regime, we have $\nu_3 T_{180^\circ} \simeq 2$, which is probably the fastest reorientation possible -- during a $90^\circ$ reorientation the \textit{Gonium} colony performs one single rotation, \textit{i.e.} each eye-spot detects the light variations only once.

A direct consequence of this dependence on light intensity is the shape of the trajectories during the reorientation process, illustrated in Fig.~\ref{fig:TurnOver-compTh}; here, only the reorienting part of the trajectories ($t > 0~\mathrm{s}$) is displayed, and all are centered at (0,0) when the light is switched on from the left. At $s_0=0.4$~lux, colonies swim a long distance before facing the light, whereas they turn much more sharply at larger $s_0$. The similar trajectories observed for $s_0 = 4$~lux and $40$~lux are consistent with saturation in phototactic response for $s_0>1$~lux seen in Fig.~\ref{fig:PhototaxEvidence}c. Note that at $s_0 = 40$~lux, a small fraction of colonies (typically 10\%) start to show erratic trajectories with a weak negative phototactic component.

A remarkable feature of  Fig.~\ref{fig:TurnOver-compTh} is that wavy trajectories gather near the center line, indicating a quick change in orientation, while smoother trajectories show a larger radius of curvature. This suggests that waviness, detrimental for swimming efficiency, is beneficial for phototactic efficiency: by providing a better scan of their environement, eye-spots from strongly unbalanced {\it Gonium} may better detect the light variations.



\subsection{\label{sec:reaction}Reaction to a step-up in light}

To explain the reorientation trajectories, we need a description of the flagella response to time-dependent variation in the light intensity. Measuring this response in freely swimming colonies is not possible because of their complex three-dimensional trajectories. To circumvent this difficulty, we use as in previous work \cite{drescher2010fidelity,leptos2018adaptive}, a micropipette technique to maintain a steady view of the colony. Because of the linearity of the Stokes flow, the fluid velocity induced by the flagellar action is proportional to the force they exert.

\begin{figure*}[t]
\includegraphics[width = 17.5cm]{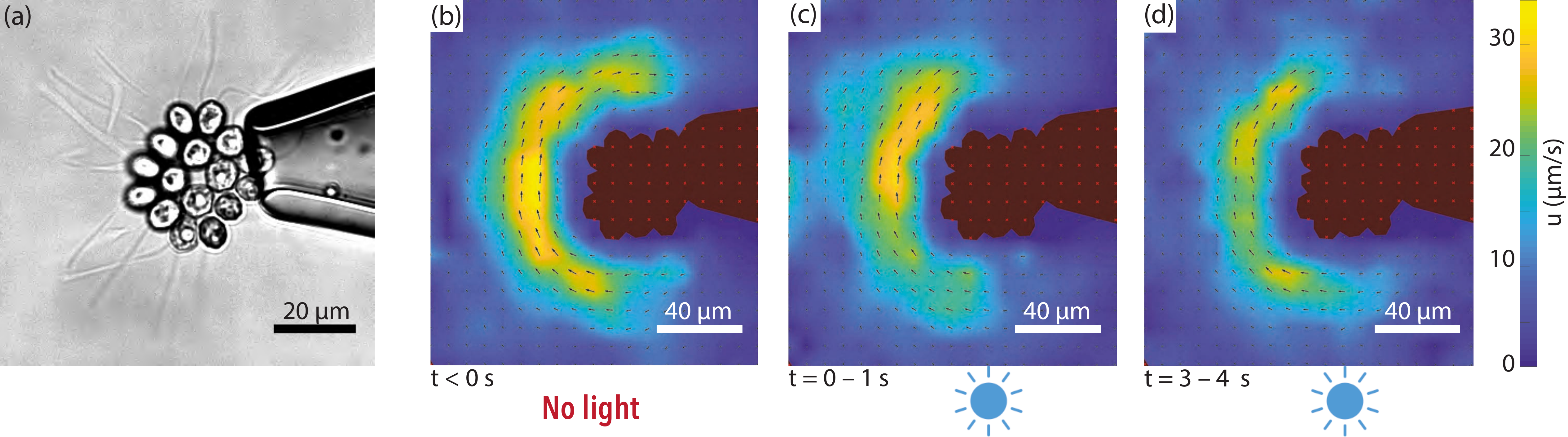}
\caption{\label{fig:pipette_circ} Micropipette experiments. 
(a) A $16$-cell colony held on a micro-pipette, viewed from posterior side. Flagella are clearly visible and their frequency can be followed as a function of time and light. The blue LED is located in the same plane as the 
micropipette, a few millimeters below the  bottom of the image.  
(b) - (d) Velocity fields measured by micro-PIV, averaged over $1$ s at three different times indicated
below each panel. The \textit{Gonium} is seen from the back (flagella away from us). Colormap is the same across the three images, from $0-30$ $\mu$m/s. Light intensity 
$s_0 \approx 1$ lux.}
\end{figure*}

The experimental setup is described in  Fig. \ref{fig:exp_setup}b and Appendix \ref{sec:app:Materials}.  Micropipettes of inner diameter slightly smaller than the \textit{Gonium} body size  (Fig. \ref{fig:pipette_circ}a) are used to catch colonies by 
gentle aspiration of fluid. An optical fiber connected to a blue LED is introduced in the chamber, and micro-PIV measurements are performed to quantify the changes in velocity field around the colony while varying the light intensity. Due to the presence of the micro-pipette, the measured velocities are not reliable on the pipette side (right-hand side of the images), but are accurate in the remainder of the field of view.

In these experiments we focus on phototactic stimulation in the form of a step-up in intensity (see also Supplementary Movie 5). The elementary flagella response to this is useful to compute the response to a more realistic change experienced by the flagella during the reorientation process. The micro-PIV experiments in Fig.~\ref{fig:pipette_circ}b-d show the flow around a \textit{Gonium} at three key moments of a step-up experiment:  prior to light excitation, the flow is nearly homogeneous and circular, with peak velocities of $30~\mathrm{\mu m/s}$ at a distance of $20-30$ 
$\mu$m from the cell body.
Immediately after light is shone from the bottom of the image (\textit{i.e.} at $90^\circ$ to the \textit{Gonium} body plane), the flow symmetry is clearly broken: the velocity is strongly reduced in the illuminated part, down to $\simeq 15~\mathrm{\mu m/s}$, about half the original value, while it remains essentially unchanged in the shadowed part. Finally, after a few seconds of constant illumination, the velocity gradually increases and the initial symmetry of the flow field is eventually recovered: the phototactic response is adaptive to the new light environment. 

\begin{figure}[h]
\includegraphics[width = 0.42\textwidth]{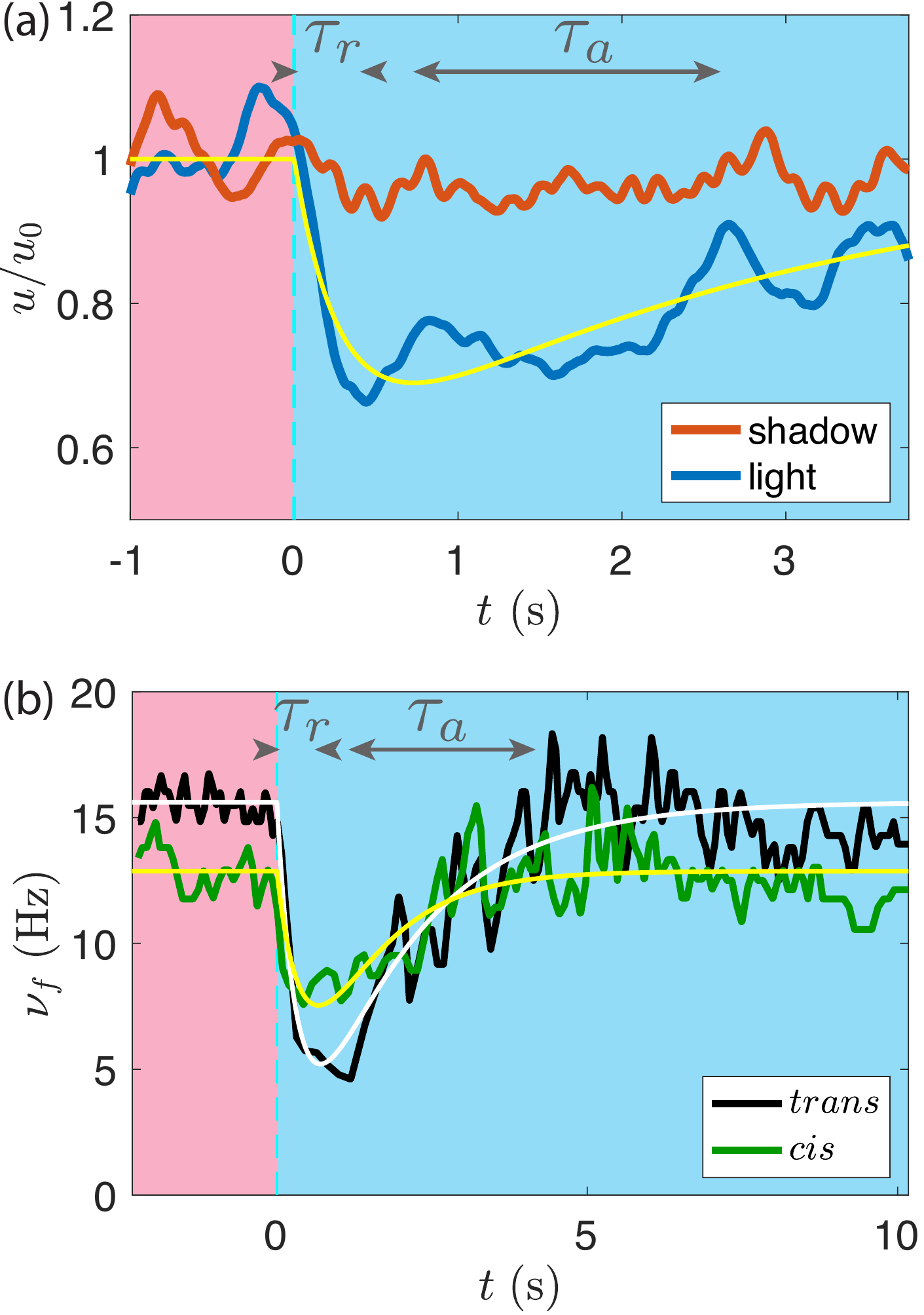}
\caption{\label{fig:flowVel} Adaptive phototactic response. (a) Normalized azimuthal velocities in 
the shaded (red curve) and illuminated (blue curve) sides of a colony held on a micropipette (averages 
over two similar experiments). For $t<0$~s there is only red illumination (no phototaxis). Blue light 
is switched on at $t=0$~s. The yellow line displays the adaptive model, Eq.~(\ref{eq:fppa2}) and (\ref{eq:pt}).
(b) $Trans$  and $cis$ flagella beat frequencies $\nu_f$ (in black and green respectively), of a single 
cell receiving light (at $s_0 \approx 1~$lux) as a function of time. White and yellow lines are the 
respective best fits with Eqs.~(\ref{eq:fppa2}) and (\ref{eq:pt}).}
\end{figure}

Flow velocities measured on the illuminated and shadowed sides are compared in Fig. \ref{fig:flowVel}a. Before light is switched on at $t=0$, both curves follow similar variations around $30~\mu$m/s. At $t=0$~s, the velocity on the illuminated side drops, while that on the opposite side remains nearly constant. After a few seconds, they tend to merge and variations are closer. As the flow velocity is proportional to the force in Stokes regime, this demonstrates a rapid drop followed by a slow recovery in the force from the illuminated side, while the force in the shadowed side remains essentially unaltered. We note that these PIV measurements only provide information on the {\it azimuthal} component of the force $f_{p\perp}$, whereas the phototactic torque (\ref{eq:lp0}) is related to a non-axisymmetric distribution of the {\it axial} component of the force $f_{p\parallel}$. 
However, because of the small tilt angle of the flagella, $\beta \simeq 30^\circ$, changes in $f_{p\parallel}$ are too difficult to detect experimentally by PIV. We assume here that the angle $\beta$ is not impacted by light, so that measurements of the azimuthal flow variations provide a good proxy for the axial flow variations. 


The velocity induced by a flagellum (hence the applied force) is a complex combination of beat frequency and waveform \cite{solari2011flagellar}. These two quantities can be measured only in simple geometries, such as in  {\it Chlamydomonas}, where the two flagella beat in the same plane. The complex three-dimensional flagellar organization in {\it Gonium} makes it difficult to quantify changes in waveform, but the response in beat frequency of each individual flagellum can be readily measured. A typical response to a step-up is displayed in Fig.~\ref{fig:flowVel}b for the two flagella ({\it cis} and {\it trans}) of a single cell detecting the light. The typical drop-and-recovery pattern of the velocity response is also remarkably present in the beating frequency, showing that the force induced by the flagella is governed at least in part by the beat frequency. The initial frequency without light stimulation is about $15$ Hz, with slightly lower values systematically found for the {\it cis}-flagellum (close to the eye-spot; see Fig.~\ref{fig:GoniumPic}b). Both flagella show a reduced beating frequency when light is switched on, with a more pronounced drop for the  {\it trans}-flagellum.  In \textit{Chlamydomonas}, the {\it cis}-flagellum shows a strong {\it decrease} \cite{witman1993, kamiya1984submicromolar} while the {\it trans}-flagellum slightly {\it increases} its frequency, a behavior which we do not observe in {\it Gonium}. Although a \textit{cis - trans} differentiation is the key to phototaxis in \textit{Chlamydomonas}, this trait, not required for phototaxis at the level of the colony, is also present at the cell level in \textit{Gonium}.

\subsection{Adaptive model}
\label{adapmodel}

Here we relate the drop-and-recovery response of the illuminated flagella to the phototactic reorientation. 
Whereas in earlier work on \textit{Volvox} phototaxis 
\cite{drescher2010fidelity} we considered the direct effect of changes in flagellar beating on the local fluid \textit{velocity} on the surface, adopting a perpsective very much like that in Lighthill's squirmer model \cite{Lighthill1952}, here we model the effect of light stimulation on the {\it force} developed by a peripheral flagella at an angle $\alpha$ as
\begin{equation}
{\bf f}_{p}(\alpha, t) = {\bf f}_{p}^{(0)} [ 1 - p_{\alpha}(t) ],
\label{eq:fppa2}
\end{equation}
with $p_\alpha$ the phototactic response. We neglect the possible phototactic response of the central flagella, which presumably do not contribute to the reorientation torque. We first describe the time-dependence of the response of a given flagellum $\alpha$ experiencing a step-up in light intensity, and then integrate the response from all flagella, taking into account eye-spot rotation, to deduce the reorientational torque, along the lines in recent
work on \textit{Chlamydomonas} \cite{leptos2018adaptive}.

The drop-and-recovery flagella response suggests using an adaptive model, which has found application in the description of sperm chemotaxis \cite{friedrich2007chemotaxis} and in phototaxis of both {\it Volvox} \cite{drescher2010fidelity} and \textit{Chlamydomonas} \cite{leptos2018adaptive}: we assume that $p(t)$ follows the light stimulation on a rapid timescale $\tau_r$, and is inhibited by an internal chemical process on a slower timescale $\tau_a$ described by a hidden variable $h(t)$. The phototactic response is assumed proportional to the light intensity, so we restrict ourselves to the linear regime at low $s_0$. The two quantities $h$ and $p$ obey a set of coupled ODEs,
\begin{subequations}
\begin{align}
\tau_r \dot p &= \mu s(t) - h - p\\
\tau_a \dot h &= \mu s(t) - h,
\end{align}
\end{subequations}
where $\mu$ is a factor with units reciprocal to those of $s(t)$. This factor represents the biological processes which link the detection of light to the subsequent physical response. In the case of a step-up in light stimulation, 
$s(t) = s_0 H(t)$, with $H$ the Heaviside function, we have
\begin{equation}
p_{step}(t) = \frac{\mu s_0}{1-\rho} [e^{-t / \tau_a} - e^{-t/\tau_r} ] H(t),
\label{eq:pt}
\end{equation}
with $\rho = \tau_r / \tau_a$. This phototactic response (\ref{eq:fppa2})-(\ref{eq:pt}), plotted as thin lines in Fig. \ref{fig:flowVel} in the case $s_0 \approx 1$~lux, provides a reasonable description of the velocity and beating frequency, with $\tau_r \simeq 0.4 \pm 0.1$~s and $\tau_a \simeq 1.5 \pm 0.5$~s.

From this fit, we infer the value of the phototactic response factor: $\mu \simeq$ 0.6 
and 0.8$\pm 0.1$~lux$^{-1}$ for the {\it cis} and {\it trans} flagella respectively. 
A somewhat lower value is obtained from the velocity signal, 
$\mu \simeq 0.4 \pm 0.1$~lux$^{-1}$, which probably results from an average over the 
set of flagella on the illuminated side and a possible influence of a change in the 
flagella beating waveform. This disparity in the evaluation of $\mu$ underlines 
the complexity of the biological processes this variable summarizes. We retain in 
the following an average value $\mu \simeq 0.6 \pm 0.2$~lux$^{-1}$.

\begin{figure}[t]
\includegraphics[width = 0.48\textwidth]{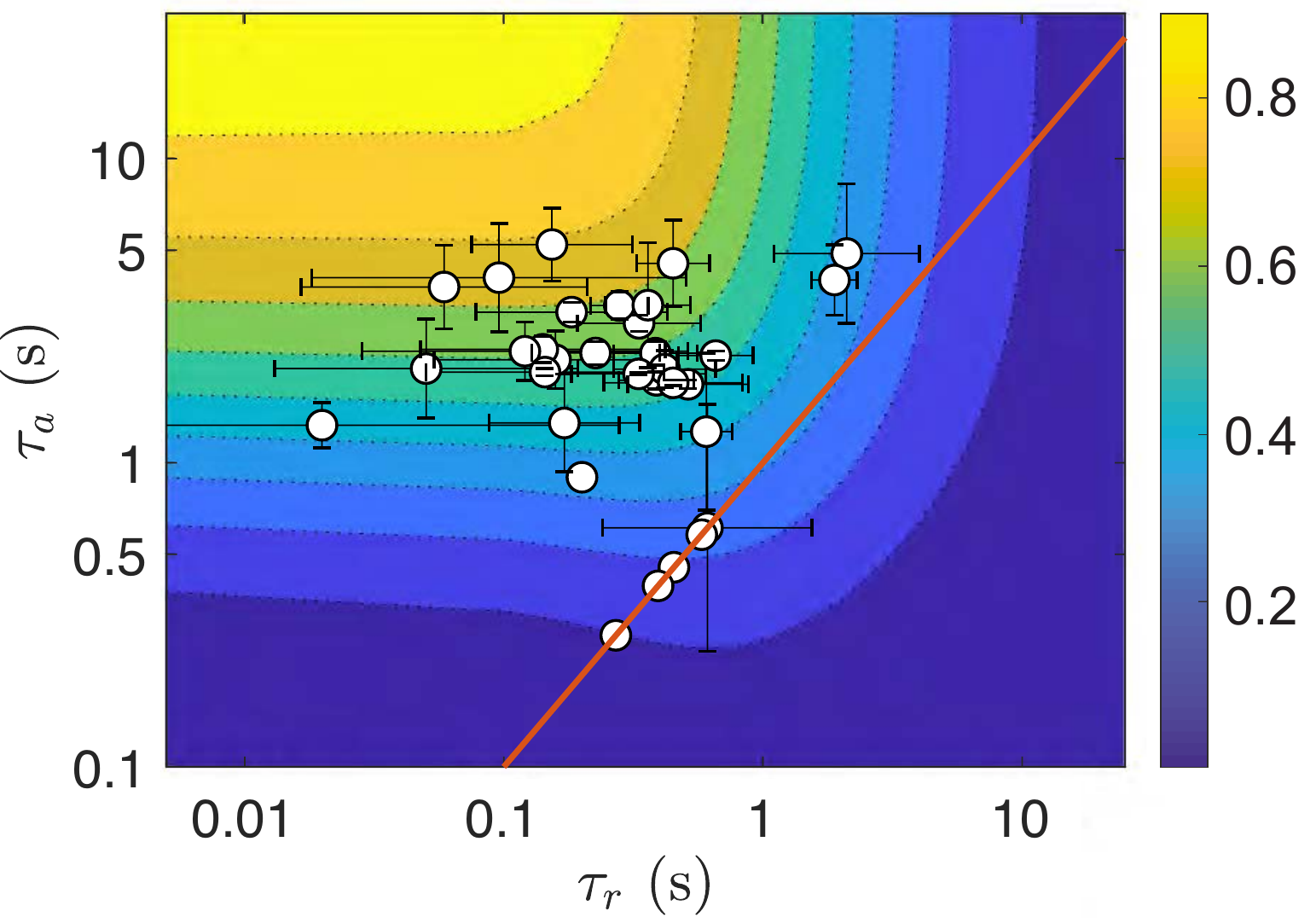}
\caption{\label{fig:gamma} Experimental characteristic times $(\tau_r, \tau_a)$ of the adaptive model.
Data obtained from the flagella beat frequencies averaged for each colony (sample size: $34$ colonies). 
Background color shows the gain function ${\cal G}$ defined in Appendix \ref{app:adap-model}, Eq.~(\ref{eq:gamma}). 
Red line indicates the relation $\tau_r = \tau_a$.}
\end{figure}

By linearity, the phototactic response to an arbitrary light stimulation $s(t)$ can be obtained as a convolution of the step-up response,
\begin{equation}
p(t) = \int_{-\infty}^t s(t') \, \frac{dp_{step}}{dt}(t-t') \, dt'.
\label{eq:rt}
\end{equation}
During the reorientation process, the light $s(t)$ perceived by the peripheral cells varies on a time scale $\omega_3^{-1}$.
According to Eq.~(\ref{eq:rt}), the response to this light variation is band-pass filtered between $\tau_a$ and $\tau_r$: an efficient response (\textit{i.e.}, a short colony reorientation time) is naturally expected for $\tau_r \ll \omega_3^{-1} \ll \tau_a$. As shown in Appendix \ref{app:optimum}, 
the maximum amplitude of the phototactic response is found in the limit $\tau_r/\tau_a \rightarrow 0$ in Eq.~(\ref{eq:pt}), yielding $p(t) = \mu s(t)$, which would correspond to a response following precisely the light stimulation.
From our data, shown in Fig. \ref{fig:gamma}, this behaviour has apparently not been selected by evolution,
and we observe large fluctuations in the characteristic times $\tau_r$ and $\tau_a$, suggesting other evolutionary advantages linked to this behaviour.
Note that in \textit{Chlamydomonas}, whose eye-spot is located at 45$^\circ$ from the $cis-$flagellum \cite{ruffer1985high}, an additional delay $\tau_d$ associated with body rotation is needed between the light detection by the eye-spot and activation of the flagellum.

To compute reorienting trajectories we consider for simplicity a perfectly balanced colony, with $\xi = 0$. In terms of phototactic reorientation, this situation is somewhat singular: such a hypothetical colony would swim in straight line, so that a change of $180^\circ$ in the light incidence, precisely opposed to their swimming direction, could not be detected by the peripheral eyespots. We therefore model a reorientation with a light incidence ${\bf \hat e}_l$ at $90^\circ$ to the initial swimming direction; this is the optimal configuration for light detection. We consider in the following ${\bf \hat e}_l = - {\bf \hat e}_x$, and an initial orientation  ${\bf \hat e}_3 = - {\bf \hat e}_y$, yielding $\theta=\pi/2$ and $\phi=0$ (see Fig.~\ref{fig:geometry}). We assume for simplicity that $\theta$ remains constant during the reorientation: the colony axis ${\bf \hat e}_3$ rotates only in the plane $({\bf \hat e}_x, {\bf \hat e}_y)$ and the phototactic torque is along ${\bf \hat e}_z$ only. At the end of the reorientation process, we have  ${\bf \hat e}_3 = {\bf \hat e}_x$, as the \textit{Gonium} is then facing the light, \textit{i.e.}  $\phi = \pi/2$.

The light perceived by a cell is
\begin{equation}
s(\kappa) = s_0 \kappa H(\kappa),
\label{eq:st}
\end{equation}
with $\kappa = - {\bf \hat e}_\ell \cdot {\bf \hat e}_r$ the projected light incidence along the local normal unit vector ${\bf \hat e}_r$,
\begin{equation}
\kappa = \cos \phi \cos (\alpha+\psi) - \cos \theta \sin \phi \sin (\alpha+\psi).
\label{eq:eler}
\end{equation}
From Eq.~(\ref{eq:rt}), at a given time $t$, each flagellum labeled by the angle $\alpha$ has a phototactic response $p_\alpha(t)$ resulting from the light $s(t')$ perceived at all times $t' < t$. This light intensity depends on its orientation, as described by Eq.~(\ref{eq:st})-(\ref{eq:eler}): it originates both from the (fast) body spin $\dot \psi = \omega_3$ and the (slow) reorientation angular velocity $\dot \phi$. 
For simplicity, the computation of $p_\alpha$ (given in Appendix \ref{app:adap-model}) assumes $\dot \psi \gg \dot \phi$, which is a valid approximation in the linear regime ($s_0<1$~lux). Using the force (\ref{eq:fppa2}), the  reorientational torque is
\begin{equation}
L_{z} = R f_{p\parallel}^{(0)} \int_{-\pi/2}^{\pi/2}  p_\alpha \, \cos \alpha \,  d \alpha.
\label{eq:lpzre}
\end{equation}
Balancing this torque with the vertical component of the frictional torque, $L_{vz} = - \eta R^3 l_1 \dot \phi$, leads to the o.d.e 
\begin{equation}
\dot \phi = \frac{1}{\tau} \cos \phi
\label{eq:ode}
\end{equation}
with a relaxation time expressed as a product of three factors, 
\begin{equation}
\tau = \frac{1}{\mu s_0}\frac{1}{{\cal G}} \frac{2 l_1\eta R^2}{\pi f_{p\parallel}^{(0)}}.
\label{eq:tau}
\end{equation}
Here, $\tau \sim 1/\mu s_0$ is a dependence consistent with the linear response assumption, and we have 
introduced a gain function ${\cal G}(\omega_3\tau_r,\omega_3 \tau_a) \in [0, 1]$ associated with the adaptive response, which is function of the non-dimensional relaxation times $\omega_3 \tau_a$ and $\omega_3 \tau_r$, and described in Appendix \ref{app:adap-model}. This gain function is such that ${\cal G} \rightarrow 1$ in the optimal case $\omega_3 \tau_r \rightarrow 0$ and $\omega_3 \tau_a \rightarrow \infty$ (no adaptive filtering), yielding the fastest reorientation time.
With the pN scale of forces, the $\sim 20-30 \mu$m scale of $R$
and the viscosity of water, we naturally find a timescale
on the order of seconds, as in experiment.

Equation~(\ref{eq:ode}) is analogous to the problem of a door pulled by a constant force with viscous friction. Its solution with initial condition $\phi(0)=0$ is
\begin{equation}
\phi(t) = \pi/2 - 2 \tan^{-1} (e^{-t/\tau}),
\label{eq:solode}
\end{equation}
which asymptotes to $\phi(\infty) = \pi/2$: we have obtained the expected phototactic reorientation, on a timescale $\tau$. In the case of a full reorientation (180$^\circ$), this solution applies only for the second half of the reorientation, for $\phi(t)$ increasing from 0 to $\pi/2$ (the first half is simply deduced by symmetrizing Eq.~(\ref{eq:solode}) for $t<0$).

\subsection{Comparison to the experiments}

\begin{figure}[t]
\includegraphics[width = 0.42\textwidth]{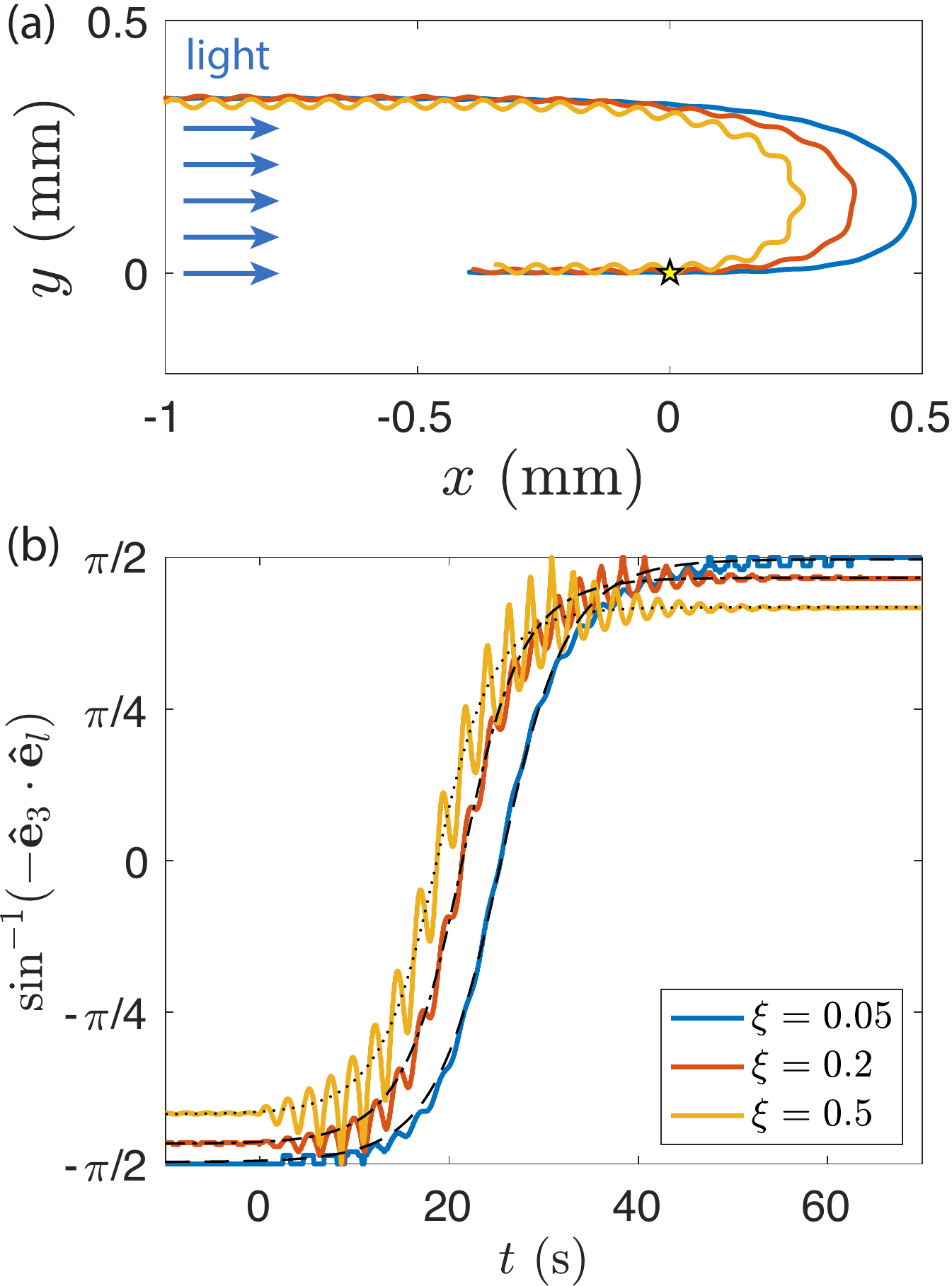}
\caption{\label{fig:num-results}Numerical results for the phototactic response. 
(a) The reorientation trajectories of three wavy swimmers are shown for $\mu s_0 = 0.5$ (corresponding to $s_0 
\approx 0.3$~lux).  They are initially swimming away from the light source, and then turn around 
after the light is switched on from the left when colonies are at the origin (yellow star).
(b) Evolution of $\phi$ as a function of time for the same swimmers. Light is turned on at $t=0$. The thin black dashed, dash-dotted and dotted lines are the respective fits using Eq.~(\ref{eq:solode}).
}
\end{figure}

To study how far {\it Gonium} colonies are from the optimum gain ${\cal G}=1$, we plot in Fig.~\ref{fig:gamma} the adaptive timescales $(\tau_r, \tau_a)$ measured for a set of colonies under illumination intensity $s_0 \approx 1$~lux.  These data were obtained by averaging the response times of the beating frequency of each flagellum. The measurements are centered around $(\tau_r, \tau_a) \approx (0.1, 2)$~s, corresponding to values of ${\cal G}$ between $0.5$ and $0.7$, which is in the upper half, but somewhat far from the optimum expected for an ideal swimmer. The error bars highlight the variability in the response times among flagella within a colony, and this evidences how biological variations divert \textit{Gonium} from its physical optimum.  

Until now, we have ignored the helical nature of the trajectories in the phototactic reorientation process.  Including waviness 
in the adaptive model of Sec.~\ref{adapmodel} would be a considerable analytical task, because the light variation perceived by 
each eyespot would depend on the three Euler angles and their time derivatives. Instead, we performed a series of direct numerical 
studies of phototactic reorientation using the computational model previously introduced, combining in the flagella force the 
azimuthal modulation (\ref{eq:def}) and the phototactic response (\ref{eq:fppa2}). Typical trajectories, obtained for various 
imbalance parameter $\xi$ (and hence pitch angle $\chi$), are illustrated in Fig.~\ref{fig:num-results}a and Supplementary Movie 3, 
for a light intensity corresponding to $\mu s_0 = 0.5$. The trajectories clearly show a faster reorientation for wavier colonies, 
consistent with the experimental observations in Fig.~\ref{fig:TurnOver-compTh}. The trajectories show about $10$ body rotations 
during the reorientation (i.e., $\dot \phi/ \dot \psi \simeq 0.05$), indicating that the linear regime assumption used in the model 
is satisfied for this value of $\mu s_0$. 

The reorientation dynamics of such wavy swimmers is illustrated in Fig.~\ref{fig:num-results}b, showing the 
angle $\sin^{-1} (- {\bf \hat e}_3 \cdot {\bf \hat e}_l)$ as a function of time. For a straight swimmer, this 
angle is simply $\phi(t)$: it increases monotonically from $-\pi/2$ to $\pi/2$ following Eq.~(\ref{eq:solode}) 
symmetrized in time. For an unbalanced swimmer, the symmetry axis ${\bf \hat e}_3$ describes a cone of apex angle 
$\zeta$ (as described in Sec.~\ref{sec:Swimming}) around the mean swimming direction: 
$\sin^{-1} (- {\bf \hat e}_3 \cdot {\bf \hat e}_l)$ therefore increases from $-\pi/2+\zeta$ to $\pi/2-\zeta$. 
The angle obtained from numerical simulations indeed shows oscillations superimposed on a mean evolution that is 
still remarkably described by the straight-swimmer law [Eq.~(\ref{eq:solode}), in dotted lines], with the asymptotic 
values $\pm \pi/2$ simply replaced by $\pm(\pi/2-\zeta)$. Increasing the imbalance parameter $\xi$ clearly increases 
the amplitude of the oscillations when the light is switched on, yielding a better scan of the environment and hence 
a shorter reorientation timescale $\tau$. Combined with the reduced mean velocity $v_m = v/\cos \chi$ (Fig.~\ref{fig:GpSwimming}c), 
this faster reorientation finally produces the sharper trajectories observed in Fig.~\ref{fig:num-results}a. Our simulations 
therefore successfully capture the main features of the phototactic reorientation dynamics observed in {\it Gonium} colonies.

\section{Conclusions}

We have presented a detailed study of motility and phototaxis of \textit{Gonium pectorale}, an organism of intermediate complexity within 
the \textit{Volvocine} green algae. In its flagellar dynamics it combines beating patterns found in \textit{Chlamydomonas} and 
\textit{Volvox}, and has a distinct symmetry compared to the
approximate bilateral symmetry of the former and the axisymmetry
of the latter.  Our experimental observations are consistent 
with a theory based on adaptive response exhibited solely by the peripheral cells, on a time scale comparable to the
rotation period of the colony around the axis normal to the 
body plane.  The precise biochemical 
pathways that underlie this adaptive response remain unclear.  
As with other green algae \cite{drescher2010fidelity,leptos2018adaptive}, the response and adaptation dynamics 
serve to define a kind of bandpass filter of response centered around
the colony rotation period, extending from tenths of a second to several seconds.  It is natural to imagine that
evolution has chosen these scales to filter out environmental fluctuations that are both very rapid (such as might occur
from undulations of the water's surface above a colony) and very slow (say, due to passing clouds), to yield 
distraction-free phototaxis.
Finally, we have observed that colonies with helical trajectories that arise from slightly imbalanced flagellar forces are 
shown to have enhanced reorientation 
dynamics relative to perfectly symmetric colonies.
Taken together, these experimental and theoretical observations lend support to a growing body of evidence \cite{jekely2009evolution} 
suggesting that helical 
swimming by tactic organisms is not only common in Nature, but possesses intrinsic biological advantages.  

\begin{acknowledgments}
We thank Kyriacos Leptos for initial experimental assistance and inspiration for the model, Lucie 
Domino for preliminary studies of \textit{Gonium}, and David Page-Croft, Caroline Kemp, and John 
Milton for vital technical support.  We are also grateful to
Matt Herron and Shota Yamashita for insights into \textit{Gonium} structure. This work was supported in part by Wellcome Trust Investigator 
Award  207510/Z/17/Z (REG \& HdM), Institut Universitaire de France (FM), 
the Japan Society for the Promotion of Science (KAKENHI Grant Nos. 17H00853 \& 17KK0080 
to TI),
Established Career Fellowship EP/M017982/1 from the Engineering 
and Physical Sciences Research Council and Grant No. 7523 from the Gordon and Betty Moore 
Foundation (REG).   
\end{acknowledgments}

\appendix

\section{\label{sec:app:Materials}Materials and methods}

\subsection*{Culture conditions}
Wild-type \textit{Gonium pectorale} colonies (strain CCAC 3275 B from the Cologne Biocenter) were 
grown in standard \textit{Volvox} medium in an incubator (Binder) at $24 ^\circ$, under 3800 lux 
illumination in a 14:10h light-dark cycle.

\subsection*{Observation of swimming}
Chambers were prepared using two glass slides, sealed with Frame-Seal Incubation Chambers (SLF0201, 
9x9~mm, BIO-RAD Laboratories). They were subsequently mounted on a Nikon TE2000-U inverted microscope for
observation with a $4\times$ or $10\times$ Nikon Plan Fluor objective.  Non-phototactic red illumination of the samples 
was achieved with a long pass filter ($620$ nm) in the light path.
Observations were recorded with a high-speed video camera (Phantom V311; see also below) at $30$ fps.
A pair of blue LEDs (M470L2, Thorlabs), connected to LED drivers (LEDD1B, Thorlabs), were 
placed on either sides of the chamber, and separately controlled, as described in Fig.~\ref{fig:exp_setup}a. 
To assure maximum photoresponse, the wavelength of 470~nm was chosen to be close to the 
absorption maximum rhodopsin \cite{hubbard1958action}, the protein in the eyespot responsible 
for light detection \cite{foster1980light}.

\subsection*{Micropipette experiments}

We used borosilicate glass pipettes (Sutter Instruments, outer diameter $1.0$ mm, inner diameter $0.75$ mm), 
a pipette puller (P-97, Sutter Instruments) and a multifunction microforge controller (DMF1000, 
World Precision Instruments) to produce micropipettes of inner diameter $\sim\!20$ $\mu$m, 
with polished edges.

Chambers were fabricated by gluing short spacers to glass slides with UV-setting glue (NOA61), with 
UV exposure for $1$ min (ELC-500 lamp, Electro-Lit Corporation), leaving apertures at $90^\circ$ for 
the micropipette and the optical fiber to enter the chamber, as sketched in Fig.~\ref{fig:exp_setup}b.
Colonies were caught on the micropipette, through which fluid was gently aspirated via a
$10$ ml syringe (BD Luer-Lok 305959). The optical fiber (FT400EMT, Thorlabs) wass connected to a 
$470$ nm LED (M470F3, Thorlabs) and driver (DC2200, Thorlabs), enabling intensity and timing control. 
We took care that the fiber end enters the liquid in the chamber to avoid losses of light by reflection 
at the air-water interface, and that it is aligned with the tip of the micropipette.  
Synchronization of the light source to the camera wass made via a NI-DAQ (BNC-2110, National Instruments).
Light intensities were measured with a lux meter (Lutron LX-101). We verified proportionality 
between the intensity (in lux) from of the optical fiber and the current (mA) provided by the driver, 
to extrapolate light intensities below $1$ lux, where the meter saturates.

Microparticle image velocimetry (PIV) experiments are conducted by adding non-fluorescent beads 
(Polybead Polystyrene Cat. 07310, Polysciences), of diameter $1~\mathrm{\mu m}$, to the 
\textit{Gonium} suspension. Image acquisition was performed on a similar microscope with a 
$20\times$ Plan Fluor (Nikon) or a $63\times$ water immersion Plan Apochromat objective (Zeiss) 
with a $45-60$ Nikon adapter, connected to the same video-camera recording at $200$ fps. Image analysis 
was performed via the Matlab tool PIVlab, with a window size adapted to the chosen microscope 
objective to ensure the presence of at least $3-5$ particles.

\section{\label{app:v3D} Inferring Helical Motion from Planar Projections}

We define three velocities for a helical trajectory: the “mean velocity” $v_m=\lambda/T$, where 
$T$ is the time to cover one wavelength $\lambda$, the true (3D) instantaneous velocity $v_{3D}$ and 
the apparent (2D) instantaneous velocity $v_{2D} = v$. From microscopy, we access 
$v_m$ and $v$, from which we deduce $v_{3D}$.
To relate these velocities, we rewrite the helical trajectory in parametric form:
\begin{equation}
\left(x(z),y(z)\right) = 
A \left(\cos k z,\sin k z\right).
\end{equation}
where $k=2\pi/\lambda$. The instantaneous velocity is $v_{3D} = l/T$, where $l=\int_0^\lambda ds$ 
is the curved length covered by \textit{Gonium}. In 3D, we have $ds = dz \sqrt{1+x_z^2 +y_z^2}$, 
yielding
\begin{align}
v_{3D} &= \frac{1}{T} \int_0^\lambda\!\! \sqrt{1+( k A)^2} dz 
\nonumber \\
&= \frac{\lambda}{T} \sqrt{1+\tan^2 \chi}
= \frac{v_m}{\cos \chi}.
\end{align}
If we assume that the 2D plane of view is $(x,z)$, then for the apparent (2D) instantaneous velocity 
$v$ we now have $ds = dz \sqrt{1+x_z^2}$, so
\begin{equation}
v_{2D} = \frac{1}{T} \int_0^\lambda \sqrt{1 + \tan^2 \chi \sin^2 (k z)}dz,
\end{equation}
which is an elliptic integral of the second kind of parameter $\tan^2 \chi$. For small $\chi$, a Taylor 
expansion gives $v_m/v_{3D} \simeq 1 - \chi^2/2$ and $v_m/v_{2D} \simeq 1-\chi^2/4$. 
For the typical values found for \textit{Gonium}, $\chi \approx 30^\circ$, we deduce that 
$v_{2D}/v_{3D} \simeq 0.93$, and therefore conclude that approximating the instantaneous velocity by 
its 2D projection is reasonable.

\section{\label{app:wavy} Helical trajectories}

We relate here the pitch angle $\chi$ of the helical trajectories to the uneven distribution 
of flagellar forces.  The geometry is shown in Fig. \ref{fig:GoniumPic}d, with ${\bf \hat e}_3$ 
the symmetry axis of the {\it Gonium} body. In this geometry, the rotation vector ${\bf \Omega}$ 
is along the mean swimming direction ${\bf \hat e}_z$. The pitch angle $\chi$ is the angle 
between the instantaneous velocity 
${\bf U}$ and ${\bf \hat e}_z$. The symmetry axis of the {\it Gonium} body describes a cone around 
$z$ of constant apex angle $\zeta$, with $\zeta<\chi$ because of the anisotropic resistance matrices. 
The Euler angles for this geometry are $\theta=\zeta$, $\dot \phi = \omega_3 / \cos \zeta$ and 
$\dot \psi = 0$ . The frictional torque (\ref{eq:lv}) is thus
\begin{equation}
{\bf L}_v = - \eta R^3
\begin{bmatrix}
l_1 \dot \phi \sin \zeta \sin \psi \\
l_1 \dot \phi \sin \zeta \cos \psi \\
l_3(\dot \psi + \dot \phi \cos \zeta)
\end{bmatrix} .
\label{eq:lvu}
\end{equation}

The simplest defect compatible with such a helical trajectory is a modulation of the axial force 
developed by the peripheral flagella in the form (c.f. Eq.~(\ref{eq:def}))
\begin{equation}
{\bf f}_{p}(\alpha) = {\bf f}_{p}^{(0)} (1+\xi \cos \alpha),
\end{equation}
with $\xi$ a ``defect'' parameter. We assume here that $\beta$ does not depend on $\alpha$, so that 
the defect affects the peripheral and axial components $f_{p \parallel}$ and $f_{p \perp}$ in the 
same way.
From this force distribution, the total force and torque (\ref{eq:lp0}) are
\begin{align}
{\bf F} &= [ F_c + 2 \pi f_{p \parallel}^{(0)} ] {\bf \hat e}_3 + \pi \xi f_{p \perp}^{(0)} 
{\bf \hat e}_2, \\
{\bf L} &= 2\pi R f_{p \perp}^{(0)} {\bf \hat e}_3 - \pi \xi R f_{p \parallel}^{(0)} 
{\bf \hat e}_2.
\end{align}
Since the force and the torque now have components along ${\bf \hat e}_2$, the velocity and 
angular velocity do as well.  The phase choice in Eq.~(\ref{eq:def}) assigns maximum 
flagella force along ${\bf \hat e}_1$,
and implies $L_1=0$, so $\psi=0$ in Eq.~(\ref{eq:lvu}).

To compute the angle $\zeta$, we first solve for the torque-angular velocity relation along 
${\bf \hat e}_2$ and ${\bf \hat e}_3$,
\begin{align}
- \eta R^3 l_1 \dot \phi \sin \zeta &= - \pi \xi R  f_{p \parallel}^{(0)}, \\
- \eta R^3 l_3 \dot \phi \cos \zeta &= 2\pi R f_{p \perp}^{(0)}.
\end{align}
Solving for $\zeta$ yields
\begin{equation}
\tan \zeta = \xi \frac{l_3 }{2 l_1 } \tan \beta
\end{equation}
with $\tan \beta = - f_{p \parallel}^{(0)} / f_{p \perp}^{(0)}$.
A symmetric \textit{Gonium} ($\xi=0$) naturally has $\zeta=0$.

The helix pitch angle $\chi$, the angle between the instantaneous velocity ${\bf U}$ 
and the mean swimming direction ${\bf \hat e}_z$, is found by 
solving the velocity-force relation 
along ${\bf \hat e}_2$ and ${\bf \hat e}_3$,
\begin{align}
\eta R k_1 U_2 &= \pi \xi f_{p \perp}^{(0)},\\
\eta R k_3 U_3 &= F_c + 2\pi f_{p \parallel}^{(0)}.
\end{align}
The angle $\zeta'$ between ${\bf \hat e}_3$ and ${\bf U}$ is
\begin{equation}
\tan \zeta'= \frac{U_2}{U_3} = \xi \frac{k_3}{2 k_1} \frac{1}{\tan \beta} 
\frac{1}{1 + F_c/(2\pi f_{p \parallel}^{(0)})}.
\end{equation}
We note that the vectors ${\bf \Omega}$ ($\parallel {\bf \hat e}_z$), ${\bf \hat e}_3$  and 
${\bf U}$ are in the same plane $({\bf \hat e}_2,{\bf \hat e}_3)$. The total angle $\chi$ between 
${\bf \hat e}_z$ and ${\bf U}$ is therefore simply $\chi = \zeta + \zeta'$. For small $\xi$, we 
have $\tan \chi \simeq \tan \zeta + \tan \zeta'$, yielding Eq.~(\ref{eq:th0}),
\begin{equation}
\tan \chi \simeq  \frac{\xi}{2} \left( \frac{l_3 }{l_1 } \tan \beta + \frac{k_3 }{k_1 } 
\frac{1}{\tan \beta} \frac{1}{1 + F_c/(2\pi f_{p \parallel}^{(0)})} \right).
\end{equation}
We can rewrite this relation using the discret point-force model. Assuming that all the 
individual flagellar forces $F_i$ are identical, we have $F_c = 8 F_{i}$ and 
$2\pi f_{p \parallel}^{(0)} = 24 F_{i} \sin \beta$, yielding 
\begin{equation}
\tan \chi \simeq \frac{\xi}{2} \left( \frac{l_3 }{l_1 } \tan \beta + \frac{k_3 }{k_1 } 
\frac{1}{\tan \beta + 1/(3\cos \beta)} \right).
\label{eq:chist}
\end{equation}

The helix parameters $A$ and $\lambda$ are derived by 
expressing the velocity in the laboratory frame, 
\begin{subequations}
\begin{align}
U_x(t)  &= - \left(U_2 \cos \zeta  - U_3 \sin \zeta\right) \sin (\dot \phi t),\\
U_y(t)  &= \left(U_2 \cos \zeta  - U_3 \sin \zeta\right)\cos (\dot \phi t),\\
U_z &=  U_2  \sin \zeta + U_3 \cos \zeta.
\end{align}
\end{subequations}
with $\dot \phi  = \omega_3 / \cos \zeta$. Here, $U_z$ corresponds to $v_m$, the mean velocity 
along $z$. Integrating these equations in time yields the helix amplitude and wavelength,
\begin{subequations}
\begin{align}
A &= \frac{U_3 \sin \zeta - U_2 \cos \zeta}{\dot \phi},\\
\lambda &= 2 \pi  \frac{U_2  \sin \zeta + U_3 \cos \zeta}{\dot \phi}.
\end{align}
\end{subequations}
Using $\tan \chi = {2 \pi A}/{\lambda}$, we obtain
\begin{equation}
\tan \chi = \frac{U_3 \tan \zeta - U_2}{U_3 + U_2 \tan \zeta},
\end{equation}
which can be shown to be equivalent to Eq.~(\ref{eq:chist}) to first order in $\xi$.

\section{\label{sec:app:Num}{Numerical methods}}

\begin{figure}[t]
\includegraphics[width = 0.48\textwidth]{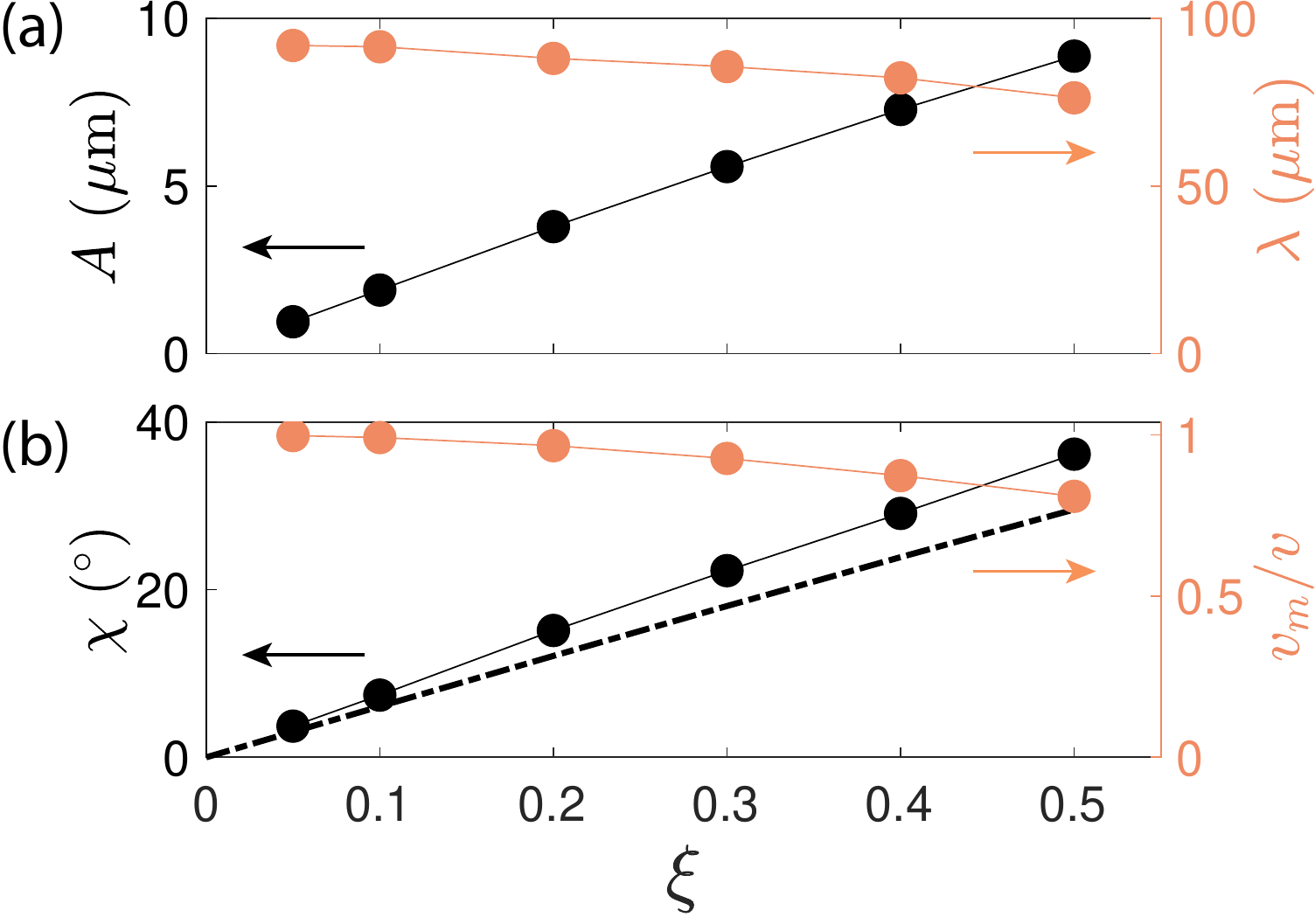}
\caption{\label{fig:app-num} Properties of wavy helical trajectories.  (a) Amplitude 
$A$ of the oscillations and wavelength $\lambda$ of the helix as a function of the defect 
parameter $\xi$. (b) Pitch angle $\chi$ and swimming efficiency $v_m/v$ as a function of the defect parameter $\xi$. The dashed line shows the theoretical prediction for $\chi$ as a function of $\xi$, obtained from Eq.~(\ref{eq:th0}).
}
\end{figure}

Here we give details of the numerical techniques and show characteristic values extracted 
from the computed wavy trajectories. We use the geometry shown in Fig. \ref{fig:geometry}b 
and described in the text. Typical Reynolds numbers for swimming \textit{Gonium} 
are $\sim 10^{-3}$, in Stokes regime. The velocity of 
the surrounding fluid at a position ${\bf x}$ can therefore be expressed as boundary 
integrals,
\begin{align}
{\bf u}({\bf x}) =& -\frac{1}{8 \pi \mu}
\int_{\substack{{\rm cell}\\{\rm body}}} \!\!\!{\bf J}({\bf x}, {\bf x'}) 
\cdot {\bf q}({\bf x'}) \, dA({\bf x'})
\nonumber \\
&-\frac{1}{8 \pi \mu} \sum_{j=1}^{32} \int_{\rm flagella}
\!\!\!\!\!\!\!\!\!\!\left[ {\bf J}({\bf x}, {\bf x'}) + {\bf W}({\bf x}, {\bf x'}) \right]
 \cdot {\bf f}({\bf x'}) \, dl_j({\bf x'})
\nonumber \\
&-\frac{1}{8 \pi \mu} \sum_{j=1}^{32}
{\bf J}({\bf x}, {\bf x'}) \cdot {\bf F}_j({\bf x'}). 
\label{vel_bem}
\end{align}
Here, ${\bf J}$ is the Stokeslet kernel, ${\bf W}$ the kernel for slender 
body theory, ${\bf q}$ the traction force on the cell body, ${\bf f}$ the traction 
force on flagella, and ${\bf F}$ the point force generated by the flagella. 
The first term of the r.h.s of Eq.~(\ref{vel_bem}) represents the drag of the cell 
body, which is solved by a boundary element method with a mesh of $320$ triangles. The second 
term accounts for the drag of the $32$ flagella, which is solved by slender body theory with 
$10$ slender elements per flagellum. The third term represents thrust and spin forces 
generated by flagella. A no-slip velocity boundary condition is applied on the cell body 
and along the flagella; force- and torque-free conditions are applied to the whole body.
Detailed functional forms and the velocity along the flagella can be found in 
Itoh \textit{et al.} \cite{Itoh2019}; here we solve Eq.~(\ref{vel_bem}) in a similar manner. 

The amplitude $A$ of the oscillations of the computed wavy trajectories, using the force 
variation around \textit{Gonium} given by Eq.~(\ref{eq:def}), linearly increases with the 
strength $\xi$ of the imbalance (Fig. \ref{fig:app-num}), while the wavelength $\lambda$ is 
hardly impacted. The deduced pitch angle $\chi$ therefore also linearly increases with 
$\xi$, in good agreement with Eq.~(\ref{eq:th0}), plotted by the dashed line. This enables us 
to deduce the experimental imbalance amplitude, as $\chi \simeq 30^\circ$ corresponds to $\xi \simeq 0.4$. Finally, the decrease in swimming efficiency $v_m/v$ follows the geometrical prediction 
$v_m/v = \cos \chi$, as plotted in Fig. \ref{fig:GpSwimming}c.

\section{\label{app:optimum} Maximum of the adaptive response}

\begin{figure}
\includegraphics[width = 0.45\textwidth]{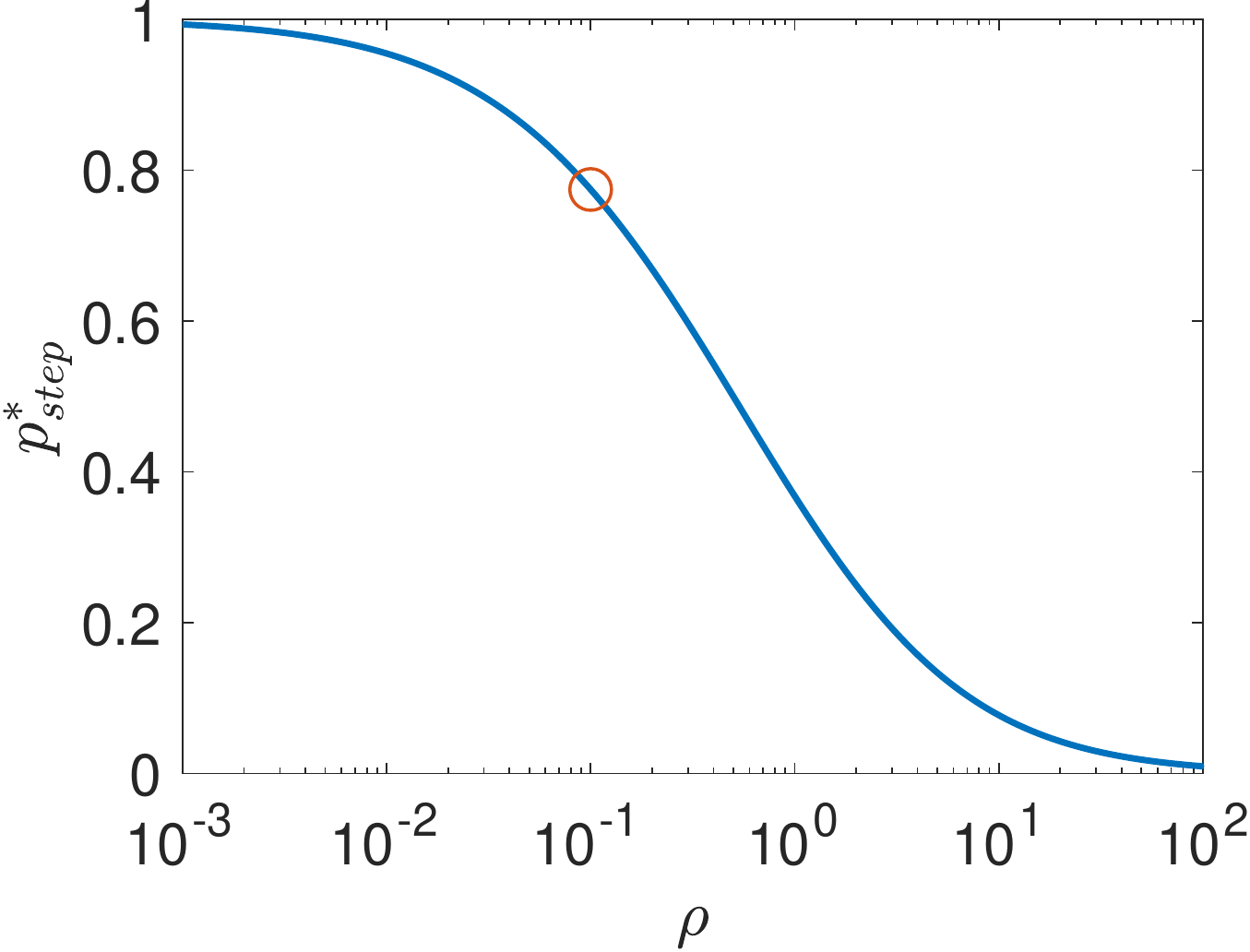}
\caption{\label{fig:stepresponse} Maximum amplitude (\ref{max_amp}) of the adaptive phototactic response to a step-up
in light intensity, as a function of the ratio $\rho=\tau_r/\tau_a$.
Red circle indicates experimental average.}
\end{figure}

Defining $s=t/\tau_r$ in the step response (\ref{eq:pt}) for $t>0$ we have
\begin{equation}
p_{step}(s) = \frac{\mu s_0}{1-\rho} \left(e^{-\rho s} - e^{-s}\right),
\label{eq:pt1}
\end{equation}
where again $\rho=\tau_r/\tau_a$. The maximum response obtained by setting $dp_{step}/ds=0$
occurs at $s^*= -(1-\rho)^{-1}\ln\rho$, with the magnitude $p^*_{step}=p_{step}(s^*)$ given by the
amusing function
\begin{equation}
p^*_{step}=\rho^{\rho/(1-\rho)}.
\label{max_amp}
\end{equation}
As shown in Fig. \ref{fig:stepresponse}, this has a maximum at $\rho=0$, confirming
the statement in Sec. \ref{adapmodel} that the maximum amplitude response is found
in the limit $\tau_r/\tau_a\to 0$.

\section{\label{app:adap-model} Phototactic gain function}

We estimate here the photoresponse $p_{\alpha}$ at an arbitrary angle $\alpha$ around 
\textit{Gonium}. This allows computation of the phototactic torque (\ref{eq:lp0}) from the 
flagella force (\ref{eq:fppa2}). The balance with the viscous torque eventually leads to the 
phototactic reorientation trajectory. This implies the definition of a gain function ${\cal G}$.
From (\ref{eq:rt}), the response to light stimulation $s(t)$ is 
\begin{equation}
p(t) = \frac{\mu}{1-\rho} \int_{-\infty}^t\!\!\! s(t') \left[\frac{e^{-(t-t')/\tau_r}}{\tau_r} - 
\frac{e^{-(t-t')/\tau_a}}{\tau_a} \right] dt'.
\label{eq:pt2}
\end{equation}
In this model, $p(t)$ has a slight negative overshoot when $s(t)$ 
decreases. However, for $\tau_r \ll \tau_a$, one has $p(t)>0$ for 
almost all time, and the 
integration of $p$ among the flagella on the illuminated side remains positive.

To evaluate the integral, we use the following geometry. We consider that only flagella on the illuminated side of the \textit{Gonium}, defined by $\alpha = -\pi/2 \cdots \pi/2$, contribute to the reorientation torque. A flagellum at angle $\alpha$ has traveled from $-\pi/2$ to $\alpha$ at constant spin $\omega_3$. After its transit to the dark side, each flagellum looses  the memory of its previous transit on the illuminated side, so that $p=0$ when it reaches $\alpha=-\pi/2$.  With $s(\alpha) = s_0 \cos \phi \cos \alpha$ and using $\alpha = \omega_3 t$, Eq.~(\ref{eq:pt2}) becomes
\begin{equation}
p_\alpha = \frac{\mu s_0 \cos \phi}{1-\rho}  \left[ I(\omega_3 \tau_r,\alpha) - I(\omega_3 \tau_a,\alpha)\right].
\label{eq:pa4}
\end{equation}
with
\begin{align}
I(A,\alpha) &= \frac{1}{A} \int_{-\pi/2}^{\alpha} \cos \alpha' \, e^{-(\alpha-\alpha')/A} \, d\alpha' \\ 
&= \frac{1}{A^2+1} \biggl[ (A \sin \alpha + \cos \alpha) \nonumber \\
&+ A e^{-(\alpha+\pi/2)/A} \biggr].
\end{align}

We finally compute the torque (\ref{eq:lp0}) from the flagella force (\ref{eq:fppa2}). Because of the delay induced by the adaptive response, the torque is no longer along ${\bf \hat e}_z$, implying a change in $\theta$. Since we assumed $\theta=\pi/2$ earlier, we consistently neglect this effect, and consider only the $z$ component of the torque,
\begin{equation}
L_{z} = R f_{p\parallel}^{(0)} \int_{-\pi/2}^{\pi/2}  p_\alpha \, \cos \alpha \,  d \alpha.
\end{equation}
Inserting (\ref{eq:pa4}), we rewrite this integral as
\begin{equation}
L_{z} = R f_{p\parallel}^{(0)} \frac{\mu s_0 \cos \phi}{1-\rho} [J(\omega_3 \tau_r) - J(\omega_3 \tau_a)],
\end{equation}
where
\begin{align}
J(A) &= \int_{-\pi/2}^{\pi/2} I(A,\alpha) \cos \alpha \, d\alpha \nonumber \\
&= \frac{1}{A^2+1} \left( \frac{\pi}{2} + \frac{A^3}{A^2+1}(1+e^{-\pi/A}) \right).
\end{align}
We have $J(0) = \pi/2$ and $J(A) \simeq 2/A$ for $A\gg1$.
Balancing this torque with the $z$ component of the friction torque $L_{vz} = - \eta R^3 l_1 \dot \phi$ yields a differential equation in the form (\ref{eq:ode}), from which we identify the gain function
\begin{equation}
{\cal G}(\omega_3\tau_r, \omega_3\tau_a) =  \frac{2}{\pi} \frac{[J(\omega_3 \tau_r) - J(\omega_3 \tau_a) ]}{1-\rho}.
\label{eq:gamma}
\end{equation}

\bibliography{BiblioGp}

\begin{thebibliography}{39}%
\makeatletter
\providecommand \@ifxundefined [1]{%
 \@ifx{#1\undefined}
}%
\providecommand \@ifnum [1]{%
 \ifnum #1\expandafter \@firstoftwo
 \else \expandafter \@secondoftwo
 \fi
}%
\providecommand \@ifx [1]{%
 \ifx #1\expandafter \@firstoftwo
 \else \expandafter \@secondoftwo
 \fi
}%
\providecommand \natexlab [1]{#1}%
\providecommand \enquote  [1]{``#1''}%
\providecommand \bibnamefont  [1]{#1}%
\providecommand \bibfnamefont [1]{#1}%
\providecommand \citenamefont [1]{#1}%
\providecommand \href@noop [0]{\@secondoftwo}%
\providecommand \href [0]{\begingroup \@sanitize@url \@href}%
\providecommand \@href[1]{\@@startlink{#1}\@@href}%
\providecommand \@@href[1]{\endgroup#1\@@endlink}%
\providecommand \@sanitize@url [0]{\catcode `\\12\catcode `\$12\catcode
  `\&12\catcode `\#12\catcode `\^12\catcode `\_12\catcode `\%12\relax}%
\providecommand \@@startlink[1]{}%
\providecommand \@@endlink[0]{}%
\providecommand \url  [0]{\begingroup\@sanitize@url \@url }%
\providecommand \@url [1]{\endgroup\@href {#1}{\urlprefix }}%
\providecommand \urlprefix  [0]{URL }%
\providecommand \Eprint [0]{\href }%
\providecommand \doibase [0]{https://doi.org/}%
\providecommand \selectlanguage [0]{\@gobble}%
\providecommand \bibinfo  [0]{\@secondoftwo}%
\providecommand \bibfield  [0]{\@secondoftwo}%
\providecommand \translation [1]{[#1]}%
\providecommand \BibitemOpen [0]{}%
\providecommand \bibitemStop [0]{}%
\providecommand \bibitemNoStop [0]{.\EOS\space}%
\providecommand \EOS [0]{\spacefactor3000\relax}%
\providecommand \BibitemShut  [1]{\csname bibitem#1\endcsname}%
\let\auto@bib@innerbib\@empty
\bibitem [{\citenamefont {Weismann}(1892)}]{Weismann1892}%
  \BibitemOpen
  \bibfield  {author} {\bibinfo {author} {\bibfnamefont {A.}~\bibnamefont
  {Weismann}},\ }\href@noop {} {\emph {\bibinfo {title} {Essays on Heredity and
  Kindred Biological Problems}}}\ (\bibinfo  {publisher} {Oxford, UK, Clarendon
  Press},\ \bibinfo {year} {1892})\BibitemShut {NoStop}%
\bibitem [{\citenamefont {Huxley}(1912)}]{Huxley1912}%
  \BibitemOpen
  \bibfield  {author} {\bibinfo {author} {\bibfnamefont {J.}~\bibnamefont
  {Huxley}},\ }\href@noop {} {\emph {\bibinfo {title} {The Individual in the
  Animal Kingdom}}}\ (\bibinfo  {publisher} {Cambridge, UK, Cambridge
  University Press},\ \bibinfo {year} {1912})\BibitemShut {NoStop}%
\bibitem [{\citenamefont {Kirk}(1998)}]{Kirkbook1998}%
  \BibitemOpen
  \bibfield  {author} {\bibinfo {author} {\bibfnamefont {D.}~\bibnamefont
  {Kirk}},\ }\href@noop {} {\emph {\bibinfo {title} {Volvox. Molecular-Genetic
  Origins of Multicellularity and Cellular Differentiation}}}\ (\bibinfo
  {publisher} {Cambridge, UK, Cambridge University Press},\ \bibinfo {year}
  {1998})\BibitemShut {NoStop}%
\bibitem [{\citenamefont {Goldstein}(2015)}]{Goldstein2015ARFM}%
  \BibitemOpen
  \bibfield  {author} {\bibinfo {author} {\bibfnamefont {R.~E.}\ \bibnamefont
  {Goldstein}},\ }\bibfield  {title} {\bibinfo {title} {Green algae as model
  organisms for biological fluid mechanics},\ }\href@noop {} {\bibfield
  {journal} {\bibinfo  {journal} {Annu. Rev. Fluid Mech.}\ }\textbf {\bibinfo
  {volume} {47}},\ \bibinfo {pages} {343} (\bibinfo {year} {2015})}\BibitemShut
  {NoStop}%
\bibitem [{\citenamefont {Goldstein}(2016)}]{Goldstein2016JFM}%
  \BibitemOpen
  \bibfield  {author} {\bibinfo {author} {\bibfnamefont {R.~E.}\ \bibnamefont
  {Goldstein}},\ }\bibfield  {title} {\bibinfo {title} {Batchelor prize
  lecture. fluid dynamics at the scale of the cell},\ }\href@noop {} {\bibfield
   {journal} {\bibinfo  {journal} {J. Fluid Mech.}\ }\textbf {\bibinfo {volume}
  {807}},\ \bibinfo {pages} {1} (\bibinfo {year} {2016})}\BibitemShut {NoStop}%
\bibitem [{\citenamefont {Hoops}(1997)}]{hoops1997motility}%
  \BibitemOpen
  \bibfield  {author} {\bibinfo {author} {\bibfnamefont {H.}~\bibnamefont
  {Hoops}},\ }\bibfield  {title} {\bibinfo {title} {Motility in the colonial
  and multicellular volvocales: structure, function, and evolution},\
  }\href@noop {} {\bibfield  {journal} {\bibinfo  {journal} {Protoplasma}\
  }\textbf {\bibinfo {volume} {199}},\ \bibinfo {pages} {99} (\bibinfo {year}
  {1997})}\BibitemShut {NoStop}%
\bibitem [{\citenamefont {Drescher}\ \emph
  {et~al.}(2010{\natexlab{a}})\citenamefont {Drescher}, \citenamefont
  {Goldstein},\ and\ \citenamefont {Tuval}}]{drescher2010fidelity}%
  \BibitemOpen
  \bibfield  {author} {\bibinfo {author} {\bibfnamefont {K.}~\bibnamefont
  {Drescher}}, \bibinfo {author} {\bibfnamefont {R.~E.}\ \bibnamefont
  {Goldstein}},\ and\ \bibinfo {author} {\bibfnamefont {I.}~\bibnamefont
  {Tuval}},\ }\bibfield  {title} {\bibinfo {title} {Fidelity of adaptive
  phototaxis},\ }\href@noop {} {\bibfield  {journal} {\bibinfo  {journal}
  {Proc. Natl. Acad. Sci. USA}\ }\textbf {\bibinfo {volume} {107}},\ \bibinfo
  {pages} {11171} (\bibinfo {year} {2010}{\natexlab{a}})}\BibitemShut {NoStop}%
\bibitem [{\citenamefont {Drescher}\ \emph
  {et~al.}(2010{\natexlab{b}})\citenamefont {Drescher}, \citenamefont
  {Goldstein}, \citenamefont {Michel}, \citenamefont {Polin},\ and\
  \citenamefont {Tuval}}]{drescher2010direct}%
  \BibitemOpen
  \bibfield  {author} {\bibinfo {author} {\bibfnamefont {K.}~\bibnamefont
  {Drescher}}, \bibinfo {author} {\bibfnamefont {R.~E.}\ \bibnamefont
  {Goldstein}}, \bibinfo {author} {\bibfnamefont {N.}~\bibnamefont {Michel}},
  \bibinfo {author} {\bibfnamefont {M.}~\bibnamefont {Polin}},\ and\ \bibinfo
  {author} {\bibfnamefont {I.}~\bibnamefont {Tuval}},\ }\bibfield  {title}
  {\bibinfo {title} {Direct measurement of the flow field around swimming
  microorganisms},\ }\href@noop {} {\bibfield  {journal} {\bibinfo  {journal}
  {Physical Review Letters}\ }\textbf {\bibinfo {volume} {105}},\ \bibinfo
  {pages} {168101} (\bibinfo {year} {2010}{\natexlab{b}})}\BibitemShut
  {NoStop}%
\bibitem [{\citenamefont {Guasto}\ \emph {et~al.}(2010)\citenamefont {Guasto},
  \citenamefont {Johnson},\ and\ \citenamefont
  {Gollub}}]{guasto2010oscillatory}%
  \BibitemOpen
  \bibfield  {author} {\bibinfo {author} {\bibfnamefont {J.~S.}\ \bibnamefont
  {Guasto}}, \bibinfo {author} {\bibfnamefont {K.~A.}\ \bibnamefont
  {Johnson}},\ and\ \bibinfo {author} {\bibfnamefont {J.~P.}\ \bibnamefont
  {Gollub}},\ }\bibfield  {title} {\bibinfo {title} {Oscillatory flows induced
  by microorganisms swimming in two dimensions},\ }\href@noop {} {\bibfield
  {journal} {\bibinfo  {journal} {Phys. Rev. Lett.}\ }\textbf {\bibinfo
  {volume} {105}},\ \bibinfo {pages} {168102} (\bibinfo {year}
  {2010})}\BibitemShut {NoStop}%
\bibitem [{\citenamefont {Bennett}\ and\ \citenamefont
  {Golestanian}(2015)}]{bennett2015steering}%
  \BibitemOpen
  \bibfield  {author} {\bibinfo {author} {\bibfnamefont {R.~R.}\ \bibnamefont
  {Bennett}}\ and\ \bibinfo {author} {\bibfnamefont {R.}~\bibnamefont
  {Golestanian}},\ }\bibfield  {title} {\bibinfo {title} {A steering mechanism
  for phototaxis in \textit{Chlamydomonas}},\ }\href@noop {} {\bibfield
  {journal} {\bibinfo  {journal} {J. R. Soc. Interface}\ }\textbf {\bibinfo
  {volume} {12}},\ \bibinfo {pages} {20141164} (\bibinfo {year}
  {2015})}\BibitemShut {NoStop}%
\bibitem [{\citenamefont {Arrieta}\ \emph {et~al.}(2017)\citenamefont
  {Arrieta}, \citenamefont {Barreira}, \citenamefont {Chioccioli},
  \citenamefont {Polin},\ and\ \citenamefont {Tuval}}]{arrieta2017phototaxis}%
  \BibitemOpen
  \bibfield  {author} {\bibinfo {author} {\bibfnamefont {J.}~\bibnamefont
  {Arrieta}}, \bibinfo {author} {\bibfnamefont {A.}~\bibnamefont {Barreira}},
  \bibinfo {author} {\bibfnamefont {M.}~\bibnamefont {Chioccioli}}, \bibinfo
  {author} {\bibfnamefont {M.}~\bibnamefont {Polin}},\ and\ \bibinfo {author}
  {\bibfnamefont {I.}~\bibnamefont {Tuval}},\ }\bibfield  {title} {\bibinfo
  {title} {Phototaxis beyond turning: persistent accumulation and response
  acclimation of the microalga \textit{Chlamydomonas reinhardtii}},\
  }\href@noop {} {\bibfield  {journal} {\bibinfo  {journal} {Sci. Rep.}\
  }\textbf {\bibinfo {volume} {7}},\ \bibinfo {pages} {3447} (\bibinfo {year}
  {2017})}\BibitemShut {NoStop}%
\bibitem [{\citenamefont {Tsang}\ \emph {et~al.}(2018)\citenamefont {Tsang},
  \citenamefont {Lam},\ and\ \citenamefont
  {Riedel-Kruse}}]{tsang2018polygonal}%
  \BibitemOpen
  \bibfield  {author} {\bibinfo {author} {\bibfnamefont {A.~C.}\ \bibnamefont
  {Tsang}}, \bibinfo {author} {\bibfnamefont {A.~T.}\ \bibnamefont {Lam}},\
  and\ \bibinfo {author} {\bibfnamefont {I.~H.}\ \bibnamefont {Riedel-Kruse}},\
  }\bibfield  {title} {\bibinfo {title} {Polygonal motion and adaptable
  phototaxis via flagellar beat switching in the microswimmer \textit{Euglena
  gracilis}},\ }\href@noop {} {\bibfield  {journal} {\bibinfo  {journal} {Nat.
  Phys.}\ }\textbf {\bibinfo {volume} {14}},\ \bibinfo {pages} {1216} (\bibinfo
  {year} {2018})}\BibitemShut {NoStop}%
\bibitem [{\citenamefont {Leptos}\ \emph {et~al.}(2018)\citenamefont {Leptos},
  \citenamefont {Chioccioli}, \citenamefont {Furlan}, \citenamefont {Pesci},\
  and\ \citenamefont {Goldstein}}]{leptos2018adaptive}%
  \BibitemOpen
  \bibfield  {author} {\bibinfo {author} {\bibfnamefont {K.~C.}\ \bibnamefont
  {Leptos}}, \bibinfo {author} {\bibfnamefont {M.}~\bibnamefont {Chioccioli}},
  \bibinfo {author} {\bibfnamefont {S.}~\bibnamefont {Furlan}}, \bibinfo
  {author} {\bibfnamefont {A.~I.}\ \bibnamefont {Pesci}},\ and\ \bibinfo
  {author} {\bibfnamefont {R.~E.}\ \bibnamefont {Goldstein}},\ }\bibfield
  {title} {\bibinfo {title} {An adaptive flagellar photoresponse determines the
  dynamics of accurate phototactic steering in \textit{Chlamydomonas}},\
  }\href@noop {} {\bibfield  {journal} {\bibinfo  {journal} {BioRxiv}\ ,\
  \bibinfo {pages} {254714}} (\bibinfo {year} {2018})}\BibitemShut {NoStop}%
\bibitem [{\citenamefont {Foster}\ and\ \citenamefont
  {Smyth}(1980)}]{foster1980light}%
  \BibitemOpen
  \bibfield  {author} {\bibinfo {author} {\bibfnamefont {K.}~\bibnamefont
  {Foster}}\ and\ \bibinfo {author} {\bibfnamefont {R.}~\bibnamefont {Smyth}},\
  }\bibfield  {title} {\bibinfo {title} {Light antennas in phototactic
  algae.},\ }\href@noop {} {\bibfield  {journal} {\bibinfo  {journal}
  {Microbiol. Rev.}\ }\textbf {\bibinfo {volume} {44}},\ \bibinfo {pages} {572}
  (\bibinfo {year} {1980})}\BibitemShut {NoStop}%
\bibitem [{\citenamefont {Hegemann}(2008)}]{hegemann2008algal}%
  \BibitemOpen
  \bibfield  {author} {\bibinfo {author} {\bibfnamefont {P.}~\bibnamefont
  {Hegemann}},\ }\bibfield  {title} {\bibinfo {title} {Algal sensory
  photoreceptors},\ }\href@noop {} {\bibfield  {journal} {\bibinfo  {journal}
  {Annu. Rev. Plant Biol.}\ }\textbf {\bibinfo {volume} {59}},\ \bibinfo
  {pages} {167} (\bibinfo {year} {2008})}\BibitemShut {NoStop}%
\bibitem [{\citenamefont {Kamiya}\ and\ \citenamefont
  {Witman}(1984)}]{kamiya1984submicromolar}%
  \BibitemOpen
  \bibfield  {author} {\bibinfo {author} {\bibfnamefont {R.}~\bibnamefont
  {Kamiya}}\ and\ \bibinfo {author} {\bibfnamefont {G.~B.}\ \bibnamefont
  {Witman}},\ }\bibfield  {title} {\bibinfo {title} {Submicromolar levels of
  calcium control the balance of beating between the two flagella in
  demembranated models of \textit{Chlamydomonas}},\ }\href@noop {} {\bibfield
  {journal} {\bibinfo  {journal} {J. Cell Bio.}\ }\textbf {\bibinfo {volume}
  {98}},\ \bibinfo {pages} {97} (\bibinfo {year} {1984})}\BibitemShut {NoStop}%
\bibitem [{\citenamefont {Josef}\ \emph {et~al.}(2005)\citenamefont {Josef},
  \citenamefont {Saranak},\ and\ \citenamefont {Foster}}]{Josef2005}%
  \BibitemOpen
  \bibfield  {author} {\bibinfo {author} {\bibfnamefont {K.}~\bibnamefont
  {Josef}}, \bibinfo {author} {\bibfnamefont {J.}~\bibnamefont {Saranak}},\
  and\ \bibinfo {author} {\bibfnamefont {K.~W.}\ \bibnamefont {Foster}},\
  }\bibfield  {title} {\bibinfo {title} {{Ciliary behavior of a negatively
  phototactic \textit{Chlamydomonas reinhardtii.}}},\ }\href
  {https://doi.org/10.1002/cm.20069} {\bibfield  {journal} {\bibinfo  {journal}
  {Cell Mot. Cytoskeleton}\ }\textbf {\bibinfo {volume} {61}},\ \bibinfo
  {pages} {97} (\bibinfo {year} {2005})}\BibitemShut {NoStop}%
\bibitem [{\citenamefont {Josef}\ \emph {et~al.}(2006)\citenamefont {Josef},
  \citenamefont {Saranak},\ and\ \citenamefont {Foster}}]{Josef2006}%
  \BibitemOpen
  \bibfield  {author} {\bibinfo {author} {\bibfnamefont {K.}~\bibnamefont
  {Josef}}, \bibinfo {author} {\bibfnamefont {J.}~\bibnamefont {Saranak}},\
  and\ \bibinfo {author} {\bibfnamefont {K.~W.}\ \bibnamefont {Foster}},\
  }\bibfield  {title} {\bibinfo {title} {{Linear systems analysis of the
  ciliary steering behavior associated with negative-phototaxis in
  \textit{Chlamydomonas reinhardtii.}}},\ }\href
  {https://doi.org/10.1002/cm.20158} {\bibfield  {journal} {\bibinfo  {journal}
  {Cell Mot. Cytoskeleton}\ }\textbf {\bibinfo {volume} {63}},\ \bibinfo
  {pages} {758} (\bibinfo {year} {2006})}\BibitemShut {NoStop}%
\bibitem [{\citenamefont {Yoshimura}\ and\ \citenamefont
  {Kamiya}(2001)}]{Yoshimura2001}%
  \BibitemOpen
  \bibfield  {author} {\bibinfo {author} {\bibfnamefont {K.}~\bibnamefont
  {Yoshimura}}\ and\ \bibinfo {author} {\bibfnamefont {R.}~\bibnamefont
  {Kamiya}},\ }\bibfield  {title} {\bibinfo {title} {{The sensitivity of
  \textit{Chlamydomonas} photoreceptor is optimized for the frequency of cell
  body rotation}},\ }\href {https://doi.org/10.1093/pcp/pce084} {\bibfield
  {journal} {\bibinfo  {journal} {Plant Cell Phys.}\ }\textbf {\bibinfo
  {volume} {42}},\ \bibinfo {pages} {665} (\bibinfo {year} {2001})}\BibitemShut
  {NoStop}%
\bibitem [{\citenamefont {Kirk}(2004)}]{kirk2004volvox}%
  \BibitemOpen
  \bibfield  {author} {\bibinfo {author} {\bibfnamefont {D.~L.}\ \bibnamefont
  {Kirk}},\ }\bibfield  {title} {\bibinfo {title} {Volvox},\ }\href@noop {}
  {\bibfield  {journal} {\bibinfo  {journal} {Curr. Bio.}\ }\textbf {\bibinfo
  {volume} {14}},\ \bibinfo {pages} {R599} (\bibinfo {year}
  {2004})}\BibitemShut {NoStop}%
\bibitem [{\citenamefont {Coleman}(2012)}]{coleman2012}%
  \BibitemOpen
  \bibfield  {author} {\bibinfo {author} {\bibfnamefont {A.~W.}\ \bibnamefont
  {Coleman}},\ }\bibfield  {title} {\bibinfo {title} {A comparative analysis of
  the volvocaceae (chlorophyta)},\ }\href@noop {} {\bibfield  {journal}
  {\bibinfo  {journal} {J. Phycol.}\ }\textbf {\bibinfo {volume} {48}},\
  \bibinfo {pages} {491} (\bibinfo {year} {2012})}\BibitemShut {NoStop}%
\bibitem [{\citenamefont {Arakaki}\ \emph {et~al.}(2013)\citenamefont
  {Arakaki}, \citenamefont {Kawai-Toyooka}, \citenamefont {Hamamura},
  \citenamefont {Higashiyama}, \citenamefont {Noga}, \citenamefont {Hirono},
  \citenamefont {Olson},\ and\ \citenamefont {Nozaki}}]{arakaki2013}%
  \BibitemOpen
  \bibfield  {author} {\bibinfo {author} {\bibfnamefont {Y.}~\bibnamefont
  {Arakaki}}, \bibinfo {author} {\bibfnamefont {H.}~\bibnamefont
  {Kawai-Toyooka}}, \bibinfo {author} {\bibfnamefont {Y.}~\bibnamefont
  {Hamamura}}, \bibinfo {author} {\bibfnamefont {T.}~\bibnamefont
  {Higashiyama}}, \bibinfo {author} {\bibfnamefont {A.}~\bibnamefont {Noga}},
  \bibinfo {author} {\bibfnamefont {M.}~\bibnamefont {Hirono}}, \bibinfo
  {author} {\bibfnamefont {B.~J.}\ \bibnamefont {Olson}},\ and\ \bibinfo
  {author} {\bibfnamefont {H.}~\bibnamefont {Nozaki}},\ }\bibfield  {title}
  {\bibinfo {title} {The simplest integrated multicellular organism unveiled},\
  }\href@noop {} {\bibfield  {journal} {\bibinfo  {journal} {PLoS One}\
  }\textbf {\bibinfo {volume} {8}},\ \bibinfo {pages} {e81641} (\bibinfo {year}
  {2013})}\BibitemShut {NoStop}%
\bibitem [{\citenamefont {Herron}\ and\ \citenamefont
  {Michod}(2008)}]{herron2008}%
  \BibitemOpen
  \bibfield  {author} {\bibinfo {author} {\bibfnamefont {M.~D.}\ \bibnamefont
  {Herron}}\ and\ \bibinfo {author} {\bibfnamefont {R.~E.}\ \bibnamefont
  {Michod}},\ }\bibfield  {title} {\bibinfo {title} {Evolution of complexity in
  the volvocine algae: transitions in individuality through darwin's eye},\
  }\href@noop {} {\bibfield  {journal} {\bibinfo  {journal} {Evolution: Int. J.
  Org. Evol.}\ }\textbf {\bibinfo {volume} {62}},\ \bibinfo {pages} {436}
  (\bibinfo {year} {2008})}\BibitemShut {NoStop}%
\bibitem [{\citenamefont {Nozaki}(1990)}]{nozaki1990ultrastructure}%
  \BibitemOpen
  \bibfield  {author} {\bibinfo {author} {\bibfnamefont {H.}~\bibnamefont
  {Nozaki}},\ }\bibfield  {title} {\bibinfo {title} {Ultrastructure of the
  extracellular matrix of \textit{Gonium} (volvocales, chlorophyta)},\
  }\href@noop {} {\bibfield  {journal} {\bibinfo  {journal} {Phycologia}\
  }\textbf {\bibinfo {volume} {29}},\ \bibinfo {pages} {1} (\bibinfo {year}
  {1990})}\BibitemShut {NoStop}%
\bibitem [{\citenamefont {Moore}(1916)}]{moore1916mechanism}%
  \BibitemOpen
  \bibfield  {author} {\bibinfo {author} {\bibfnamefont {A.}~\bibnamefont
  {Moore}},\ }\bibfield  {title} {\bibinfo {title} {The mechanism of
  orientation in \textit{Gonium}},\ }\href@noop {} {\bibfield  {journal}
  {\bibinfo  {journal} {J. Exp. Zool.}\ }\textbf {\bibinfo {volume} {21}},\
  \bibinfo {pages} {431} (\bibinfo {year} {1916})}\BibitemShut {NoStop}%
\bibitem [{\citenamefont {Mast}(1916)}]{mast1916process}%
  \BibitemOpen
  \bibfield  {author} {\bibinfo {author} {\bibfnamefont {S.~O.}\ \bibnamefont
  {Mast}},\ }\bibfield  {title} {\bibinfo {title} {The process of orientation
  in the colonial organism, \textit{Gonium pectorale}, and a study of the
  structure and function of the eye-spot},\ }\href@noop {} {\bibfield
  {journal} {\bibinfo  {journal} {J. Exp. Zool.}\ }\textbf {\bibinfo {volume}
  {20}},\ \bibinfo {pages} {1} (\bibinfo {year} {1916})}\BibitemShut {NoStop}%
\bibitem [{\citenamefont {Greuel}\ and\ \citenamefont
  {Floyd}(1985)}]{greuel1985}%
  \BibitemOpen
  \bibfield  {author} {\bibinfo {author} {\bibfnamefont {B.~T.}\ \bibnamefont
  {Greuel}}\ and\ \bibinfo {author} {\bibfnamefont {G.~L.}\ \bibnamefont
  {Floyd}},\ }\bibfield  {title} {\bibinfo {title} {Development of the
  flagellar apparatus and flagellar orientation in the colonial green alga
  \textit{Gonium pectorale} (volvocales)},\ }\href@noop {} {\bibfield
  {journal} {\bibinfo  {journal} {J. Phycol.}\ }\textbf {\bibinfo {volume}
  {21}},\ \bibinfo {pages} {358} (\bibinfo {year} {1985})}\BibitemShut
  {NoStop}%
\bibitem [{\citenamefont {Harper}(1912)}]{harper1912}%
  \BibitemOpen
  \bibfield  {author} {\bibinfo {author} {\bibfnamefont {R.~A.}\ \bibnamefont
  {Harper}},\ }\bibfield  {title} {\bibinfo {title} {The structure and
  development of the colony in \textit{Gonium}},\ }\href@noop {} {\bibfield
  {journal} {\bibinfo  {journal} {Trans. Am. Micro. Soc.}\ }\textbf {\bibinfo
  {volume} {31}},\ \bibinfo {pages} {65} (\bibinfo {year} {1912})}\BibitemShut
  {NoStop}%
\bibitem [{\citenamefont {M{\"u}ller}(1782)}]{muller1782}%
  \BibitemOpen
  \bibfield  {author} {\bibinfo {author} {\bibfnamefont {O.~F.}\ \bibnamefont
  {M{\"u}ller}},\ }\href@noop {} {\emph {\bibinfo {title} {Kleine Schriften aus
  der Naturhistorie. p. 15-21}}}\ (\bibinfo  {publisher} {Dessau, herausgegeben
  von JAE Goeze},\ \bibinfo {year} {1782})\BibitemShut {NoStop}%
\bibitem [{\citenamefont {Witman}(1993)}]{witman1993}%
  \BibitemOpen
  \bibfield  {author} {\bibinfo {author} {\bibfnamefont {G.~B.}\ \bibnamefont
  {Witman}},\ }\bibfield  {title} {\bibinfo {title} {\textit{Chlamydomonas}
  phototaxis},\ }\href@noop {} {\bibfield  {journal} {\bibinfo  {journal}
  {Trends cell biol.}\ }\textbf {\bibinfo {volume} {3}},\ \bibinfo {pages}
  {403} (\bibinfo {year} {1993})}\BibitemShut {NoStop}%
\bibitem [{\citenamefont {Lauga}\ and\ \citenamefont
  {Powers}(2009)}]{lauga2009hydrodynamics}%
  \BibitemOpen
  \bibfield  {author} {\bibinfo {author} {\bibfnamefont {E.}~\bibnamefont
  {Lauga}}\ and\ \bibinfo {author} {\bibfnamefont {T.~R.}\ \bibnamefont
  {Powers}},\ }\bibfield  {title} {\bibinfo {title} {The hydrodynamics of
  swimming microorganisms},\ }\href@noop {} {\bibfield  {journal} {\bibinfo
  {journal} {Rep. Prog. Phys.}\ }\textbf {\bibinfo {volume} {72}},\ \bibinfo
  {pages} {096601} (\bibinfo {year} {2009})}\BibitemShut {NoStop}%
\bibitem [{\citenamefont {Symon}(1971)}]{symon1971chapter}%
  \BibitemOpen
  \bibfield  {author} {\bibinfo {author} {\bibfnamefont {K.~R.}\ \bibnamefont
  {Symon}},\ }\href@noop {} {\emph {\bibinfo {title} {Mechanics}}},\ \bibinfo
  {edition} {3rd}\ ed.\ (\bibinfo  {publisher} {Reading, MA, Addison-Wesley},\
  \bibinfo {year} {1971})\ Chap.~\bibinfo {chapter} {11}\BibitemShut {NoStop}%
\bibitem [{\citenamefont {Ito}\ \emph {et~al.}(2019)\citenamefont {Ito},
  \citenamefont {Omori},\ and\ \citenamefont {Ishikawa}}]{Itoh2019}%
  \BibitemOpen
  \bibfield  {author} {\bibinfo {author} {\bibfnamefont {H.}~\bibnamefont
  {Ito}}, \bibinfo {author} {\bibfnamefont {T.}~\bibnamefont {Omori}},\ and\
  \bibinfo {author} {\bibfnamefont {T.}~\bibnamefont {Ishikawa}},\ }\bibfield
  {title} {\bibinfo {title} {Swimming mediated by ciliary beating: comparison
  with a squirmer model},\ }\href@noop {} {\bibfield  {journal} {\bibinfo
  {journal} {J. Fluid Mech.}\ }\textbf {\bibinfo {volume} {874}},\ \bibinfo
  {pages} {774} (\bibinfo {year} {2019})}\BibitemShut {NoStop}%
\bibitem [{\citenamefont {Solari}\ \emph {et~al.}(2011)\citenamefont {Solari},
  \citenamefont {Drescher}, \citenamefont {Ganguly}, \citenamefont {Kessler},
  \citenamefont {Michod},\ and\ \citenamefont
  {Goldstein}}]{solari2011flagellar}%
  \BibitemOpen
  \bibfield  {author} {\bibinfo {author} {\bibfnamefont {C.~A.}\ \bibnamefont
  {Solari}}, \bibinfo {author} {\bibfnamefont {K.}~\bibnamefont {Drescher}},
  \bibinfo {author} {\bibfnamefont {S.}~\bibnamefont {Ganguly}}, \bibinfo
  {author} {\bibfnamefont {J.~O.}\ \bibnamefont {Kessler}}, \bibinfo {author}
  {\bibfnamefont {R.~E.}\ \bibnamefont {Michod}},\ and\ \bibinfo {author}
  {\bibfnamefont {R.~E.}\ \bibnamefont {Goldstein}},\ }\bibfield  {title}
  {\bibinfo {title} {Flagellar phenotypic plasticity in volvocalean algae
  correlates with p{\'e}clet number},\ }\href@noop {} {\bibfield  {journal}
  {\bibinfo  {journal} {J. R. Soc. Interface}\ }\textbf {\bibinfo {volume}
  {8}},\ \bibinfo {pages} {1409} (\bibinfo {year} {2011})}\BibitemShut
  {NoStop}%
\bibitem [{\citenamefont {Lighthill}(1952)}]{Lighthill1952}%
  \BibitemOpen
  \bibfield  {author} {\bibinfo {author} {\bibfnamefont {M.~J.}\ \bibnamefont
  {Lighthill}},\ }\bibfield  {title} {\bibinfo {title} {{On the squirming
  motion of nearly spherical deformable bodies through liquids at very small
  Reynolds numbers}},\ }\href@noop {} {\bibfield  {journal} {\bibinfo
  {journal} {Comm. Pure Appl. Math.}\ }\textbf {\bibinfo {volume} {5}},\
  \bibinfo {pages} {109} (\bibinfo {year} {1952})}\BibitemShut {NoStop}%
\bibitem [{\citenamefont {Friedrich}\ and\ \citenamefont
  {J{\"u}licher}(2007)}]{friedrich2007chemotaxis}%
  \BibitemOpen
  \bibfield  {author} {\bibinfo {author} {\bibfnamefont {B.~M.}\ \bibnamefont
  {Friedrich}}\ and\ \bibinfo {author} {\bibfnamefont {F.}~\bibnamefont
  {J{\"u}licher}},\ }\bibfield  {title} {\bibinfo {title} {Chemotaxis of sperm
  cells},\ }\href@noop {} {\bibfield  {journal} {\bibinfo  {journal} {Proc.
  Natl. Acad. Sci. USA}\ }\textbf {\bibinfo {volume} {104}},\ \bibinfo {pages}
  {13256} (\bibinfo {year} {2007})}\BibitemShut {NoStop}%
\bibitem [{\citenamefont {R{\"u}ffer}\ and\ \citenamefont
  {Nultsch}(1985)}]{ruffer1985high}%
  \BibitemOpen
  \bibfield  {author} {\bibinfo {author} {\bibfnamefont {U.}~\bibnamefont
  {R{\"u}ffer}}\ and\ \bibinfo {author} {\bibfnamefont {W.}~\bibnamefont
  {Nultsch}},\ }\bibfield  {title} {\bibinfo {title} {High-speed
  cinematographic analysis of the movement of \textit{Chlamydomonas}},\
  }\href@noop {} {\bibfield  {journal} {\bibinfo  {journal} {Cell Mot.}\
  }\textbf {\bibinfo {volume} {5}},\ \bibinfo {pages} {251} (\bibinfo {year}
  {1985})}\BibitemShut {NoStop}%
\bibitem [{\citenamefont {J{\'e}kely}(2009)}]{jekely2009evolution}%
  \BibitemOpen
  \bibfield  {author} {\bibinfo {author} {\bibfnamefont {G.}~\bibnamefont
  {J{\'e}kely}},\ }\bibfield  {title} {\bibinfo {title} {Evolution of
  phototaxis},\ }\href@noop {} {\bibfield  {journal} {\bibinfo  {journal}
  {Phil. Trans. R. Soc. B}\ }\textbf {\bibinfo {volume} {364}},\ \bibinfo
  {pages} {2795} (\bibinfo {year} {2009})}\BibitemShut {NoStop}%
\bibitem [{\citenamefont {Hubbard}\ and\ \citenamefont
  {Kropf}(1958)}]{hubbard1958action}%
  \BibitemOpen
  \bibfield  {author} {\bibinfo {author} {\bibfnamefont {R.}~\bibnamefont
  {Hubbard}}\ and\ \bibinfo {author} {\bibfnamefont {A.}~\bibnamefont
  {Kropf}},\ }\bibfield  {title} {\bibinfo {title} {The action of light on
  rhodopsin},\ }\href@noop {} {\bibfield  {journal} {\bibinfo  {journal} {Proc.
  Natl. Acad. Sci. USA}\ }\textbf {\bibinfo {volume} {44}},\ \bibinfo {pages}
  {130} (\bibinfo {year} {1958})}\BibitemShut {NoStop}%
\end{thebibliography}%

\end{document}